\documentclass[12pt, authoryear]{elsarticle}
\usepackage[english]{babel}
\usepackage{a4wide,amssymb}%
\usepackage{epsfig,epstopdf,graphicx,float,color,subfigure}
\usepackage[tbtags]{amsmath}
\usepackage{setspace}
\usepackage{rotating}
\usepackage{pdflscape}
\usepackage{caption}
\usepackage{multirow}

\def\rr#1{(\ref{#1})}
\def\bm#1{\mbox{\boldmath{$#1$}}}

\newcommand{\ep}{\omega}

\newcommand{\paa}{\partial}
\def\ii{{\rm i}}
\newcommand{\s}[1]{{\Large\textsf{\textbf{#1}}}}
\usepackage{lineno}
\allowdisplaybreaks

\begin{document}
\title{\s{Wrinkling of differentially growing bilayers with similar film and substrate moduli}}
\author[add1,add3]{Jiajia Shen}
\ead{j.shen3@exeter.ac.uk}
\author[add2]{Yibin Fu\corref{cor1}
}
\ead{y.fu@keele.ac.uk}
\author[add1]{Alberto Pirrera}
\ead{alberto.pirrera@bristol.ac.uk}
\author[add1]{Rainer M.~J.~Groh}
\ead{rainer.groh@bristol.ac.uk}

\cortext[cor1]{Corresponding author}

\address[add1]{Bristol Composites Institute (BCI),  School of Civil, Aerospace and Design Engineering, \\ University of Bristol, Bristol BS8 1TR, UK}
\address[add3]{Exeter Technologies Group (ETG), Department of Engineering, \\ University of Exeter, Exeter EX4 4QF, UK}
\address[add2]{School of Computing and Mathematics, Keele University, Staffs ST5 5BG, UK}

\begin{keyword}
Generalised path-following;
Nonlinear elasticity;
Stability and bifurcation; Multi-stability;
Asymptotic analysis;
Morphoelasticity.
\end{keyword}

\begin{frontmatter}
\begin{abstract}
 The study of growth-induced surface wrinkling in constrained bilayers comprising a thin film attached to a thick substrate is a canonical model for understanding pattern formation in many biological systems. While the bilayer model has received much prior attention, the nonlinear behaviour for arrangements with similar film and substrate properties, or substrate growth that outpaces film growth, remains poorly understood. This paper therefore focuses on these cases in which the substrate's elasticity dominates surface wrinkling.
We study the critical states, i.e.\ the equilibrium configurations at the onset of wrinkling, and the initial and advanced post-critical behaviour of growing bilayers with film-to-substrate modulus ratios ($\mu_\mathrm{f}/\mu_\mathrm{s}$) in the region of $2.5$--$50$, and cases where the substrate grows faster than the film, i.e.\ differential growth ratios ($g_\mathrm{s}/g_\mathrm{f}$) greater than one.
 Based on nonlinear elasticity, we formulate analytical models for linear buckling analyses and asymptotic projections around the critical point, and use finite element (FE) models coupled to continuation and branch-switching algorithms to uncover the deep post-critical regime.
 For the critical state, both exact and asymptotic solutions are derived, which are used to understand the role that the film/substrate modulus and differential growth ratios play in influencing the critical strain, critical wavelength and initial post-wrinkling behaviour.
 We demonstrate that the widely-accepted first-order asymptotic expression for the critical strain corresponding to uniform isotropic, constrained growth loses accuracy when the film/substrate modulus ratio is less than 50, and provide higher order corrections.
 It is shown that a rapidly growing substrate changes the critical mode from film-governed sinusoidal wrinkling to substrate-governed Biot wrinkling and this transition is a function of both $\mu_\mathrm{f}/\mu_\mathrm{s}$ and $g_\mathrm{s}/g_\mathrm{f}$. Interestingly, the transition from super- to subcritical bifurcation does not always coincide with the onset of a Biot mode, but rather forms a separate boundary in the $\mu_\mathrm{f}/\mu_\mathrm{s}$--$g_\mathrm{s}/g_\mathrm{f}$ space, i.e.\ unstable sinusoidal wrinkling is possible and occurs separately from the transition to Biot wrinkling. Finally, we present a phase change diagram of the post-critical modal landscape split into sinusoidal wrinkling, period doubling, period quadrupling, and creasing regimes in terms of $\mu_\mathrm{f}/\mu_\mathrm{s}$--$g_\mathrm{s}/g_\mathrm{f}$. While the post-critical regime of film- and substrate-dominated bilayers (either in terms of dominant elasticity or growth rate) is governed by sinusoidal wrinkling and Biot creasing, respectively, the intermediate regions allow for period doubling and quadrupling bifurcations. Finally, we demonstrate the existence of multi-stability in the advanced post-buckling regimes for growing bilayers where growth in the substrate surpasses that of the film.
\end{abstract}
\end{frontmatter}

\section{Introduction}
Growth-induced morphological instabilities are ubiquitously observed in biological systems across various length scales, owing to the intrinsically low elastic moduli and mechanical constraints imposed by tissues growing at different rates \citep{li2012mechanics,WANG2024105534}. Indeed, the ubiquity of wrinkling, folding and creasing means that these instabilities are closely correlated to healthy tissue function and could serve as clinical indicators of pathological conditions \citep{harris2004gyrification,KUHL2014Review,ciarletta2014pattern}. A common idealised material arrangement consists of a thin stiffer layer mounted on top of a thicker, more compliant substrate, where either the thin outer layer, the substrate, or both grow at specific rates. Depending on the film/substrate stiffness ratio $\mu_\mathrm{f}/\mu_\mathrm{s}$, growth ratio $g_\mathrm{f}/g_\mathrm{s}$, and the overall magnitude of growth (which we take to be $g_\mathrm{f}$ for convenience), different morphological patterns such as sinusoidal wrinkling, period doubling, period quadrupling, folding or creasing are observed \citep{wang2015GrowthSurfaceBuckling,Razavi2016}. In this regard, a morphological phase change diagram~\citep{wang2015GrowthSurfaceBuckling} of growing bilayers in terms of  $\mu_\mathrm{f}/\mu_\mathrm{s}$,  $g_\mathrm{f}/g_\mathrm{s}$ and  $g_\mathrm{f}$ is an effective way of describing possible wrinkling morphologies.
This work revisits the now classical problem of differentially growing bilayers to address hitherto unsolved questions in arrangements where $\mu_\mathrm{f}/\mu_\mathrm{s}< 50$ and $g_\mathrm{s}/g_\mathrm{f} > 1$ by using a combination of analytical and numerical methods.


 The critical behaviour and morphological pattern formation of growing bilayers constrained at their ends with large $\mu_\mathrm{f}/\mu_\mathrm{s}$ has been investigated using both analytical~\citep{Li2011,alawiye2019revisiting,Alawiye2020} and numerical approaches~\citep{wang2014phase,liu2017robust,jin2019post,Dortdivanlioglu2017}. Generally speaking, the most commonly analysed system is one where only the film grows or both film and substrate grow at the same rate (uniform growth). In these cases, the post-critical response is stable and sequential mode progression occurs from sinusoidal wrinkling to period doubling and potentially further doubling events until self-contact occurs. The doubling events reflect the tendency of stiffer external layers to localise into the substrate when the substrate elasticity softens with increasing compression~\citep{brau2011multiple} or increasing constrained growth.

The nonlinearity of the substrate can be accentuated by embedding pre-stretch (pre-compression or pre-tension) into the substrate before the film is attached, thereby increasing the complexity of the wrinkling behaviour observed~\citep{hutchinson2013role}. In the case of pre-tension, outward ridges may form~\citep{JIN2015Wrinkle}, while for pre-compression seemingly random wrinkling modes are possible~\citep{auguste2014PostWringBifur,Shen2022ProgramBilayer} or wrinkling is replaced entirely by the subcritical formation of a self-contacting crease~\citep{wang2015GrowthSurfaceBuckling}, reminiscent of the surface instability of a compressed solid block \citep{Gent1999, Hong2009,jin2011creases,hohlfeld2011unfolding,Jin2015,CT2019, yang2021perturbation, pandurangi2022nucleation}. In the case of growing bilayers, effective pre-stretch occurs naturally if the film and substrate grow at different rates, and as elucidated above, the case when the substrate's nonlinear effect is pronounced, leads to particularly rich behaviour. This is the case when the substrate grows faster than the film---a scenario which is seldom considered in the literature.


Another open question is how the critical wrinkling mode of an end-constrained and growing bilayer evolves as the film and substrate elastic properties start to converge ($\mu_\mathrm{f}/\mu_\mathrm{s} \rightarrow 1$). In this scenario, the bilayer starts to approach the compressed hyperelastic halfspace addressed by \cite{biot1963surface}, owing to the fact that the bulk of the system---i.e.\ the substrate---dominates the mechanical response. It is well known that surface wrinkling of an axially compressed hyperelastic halfspace is a subcritical bifurcation event, but it remains an open question for which combinations of $\mu_\mathrm{f}/\mu_\mathrm{s}$ and $g_\mathrm{s}/g_\mathrm{f}$ the transition from super- to subcritical wrinkling occurs for a bilayer and if this transition is coincident with the onset of an infinite-wavelength Biot mode or rather occurs within the regime of sinusoidal wrinkling. This paper intends to address this question and, in doing so, also assists in shedding light on crease formation in an uniaxially compressed hyperelastic halfspace.

Beyond the critical strain, relatively little work has been conducted on tracing out the post-critical bifurcation landscape, especially for the combination of elastic properties ($\mu_\mathrm{f}/\mu_\mathrm{s}<50$) and differential growth ratios ($g_\mathrm{s}/g_\mathrm{f}>1$) considered here. For example, \cite{wang2015GrowthSurfaceBuckling} treated the bilayer as a thermodynamic system that seeks to globally minimise its total potential energy and thereby constructed a post-critical phase diagram based on a Maxwell energy criterion. An initial mismatch strain is first embedded between film and substrate and only the film is then compressed, thereby modelling constrained growth of the film with a pre-stretched substrate. Our goal here is to use numerical continuation techniques~\citep{groh2022morphoelastic} to explore the post-critical bifurcation landscape and determine a three-dimensional phase diagram that summarises the post-critical wrinkling evolution as a function of $\mu_\mathrm{f}/\mu_\mathrm{s}$ and $g_\mathrm{s}/g_\mathrm{f}$, i.e.\ both the film and substrate grow from an unstrained initial state, but at different rates.

In summary, this paper describes the mechanics that govern the nonlinear behaviour of growing bilayers with film-to-substrate modulus ratios within the range seen in most biological systems and for cases where the substrate grows faster than the film. In our approach, we use both analytical bifurcation methods and finite element modeling with numerical continuation and branch switching~\citep{groh2022morphoelastic}, and provide a complete picture concerning the bifurcation behaviour in the linear, weakly nonlinear, and fully nonlinear regimes. The main contributions and findings of the paper are:
\begin{itemize}
    \item In view of the fact that the commonly employed leading-order asymptotic expansion (with $\mu_\mathrm{s}/\mu_\mathrm{f}$ taken as a small parameter) of the critical growth magnitude for uniform film/substrate growth loses accuracy for $\mu_\mathrm{f}/\mu_\mathrm{s}<50$, we provide higher-order corrections that significantly improve the accuracy of the asymptotic expansion.
    \item A curve in the $\mu_\mathrm{f}/\mu_\mathrm{s}$--$g_\mathrm{s}/g_\mathrm{f}$ plane that marks the super-critical to sub-critical transition of the nature of the first bifurcation is determined (extending the classical result for uniaxial compression that the first bifurcation is super-critical if $\mu_\mathrm{f}/\mu_\mathrm{s}>1.75$ and sub-critical if $\mu_\mathrm{f}/\mu_\mathrm{s}<1.75$). Also, it is shown that at this transition curve/boundary, the critical mode corresponds to sinusoidal wrinkling (non-zero wavenumber) if $\mu_\mathrm{f}/\mu_\mathrm{s}<6.31$ but to a Biot mode (zero wavenumber) if $\mu_\mathrm{f}/\mu_\mathrm{s}>6.31$.
    \item Phase change diagrams in the $\mu_\mathrm{f}/\mu_\mathrm{s}$--$g_\mathrm{s}/g_\mathrm{f}$--$g_\mathrm{f}$ space are constructed that split the modal landscape into sinusoidal wrinkling, period doubling, period quadrupling, and creasing regimes.
    \item When substrate growth dominates, long bilayers can wrinkle into many different modes (multi-stability) that can be constructed from  a finite set of wrinkling patterns.
\end{itemize}


The rest of the paper is structured as follows. Section 2 describes the analytical formulation in terms of a two-dimensional (2D) plane strain, compressible and hyperelastic model with independent film and substrate growth. The FE model and numerical continuation algorithms are also presented.  Sections 3 and 4 are devoted to the derivation of the exact bifurcation condition and a weakly nonlinear near-critical analysis, respectively. The bifurcation landscape and phase change diagrams are then discussed in Section 5 with the aid of numerical simulations. Finally, concluding remarks are made in Section 6.
\section{Theory}
\label{Sec:theory}

\subsection{Problem formulation}
We first present the governing equations for a general homogeneous elastic body $B$ which is composed of a non-heat-conducting compressible elastic material. The elastic body $B$ is assumed to possess an initial unstressed state $B_\mathrm{0}$.  Due to restricted growth, a purely homogeneous static deformation takes place and brings $ B_\mathrm{0}$ to a finitely stressed equilibrium configuration denoted by $ B_\mathrm{e}$. In order to search for other bifurcated configurations, we superimpose on $B_\mathrm{e}$ a small amplitude perturbation, and the resulting configuration, termed the current configuration, is denoted by $B_\mathrm{t}$. The components of the position vectors of a representative material particle relative to a common rectangular coordinate system are denoted by $X_\mathrm{A}, x_{i} (X_\mathrm{A})$ and $\tilde{x}_{i} (X_\mathrm{A})$ in $B_\mathrm{0}, B_\mathrm{e}$ and $B_\mathrm{t}$ respectively. We write
\begin{equation}
\begin{aligned}
\tilde{x}_{i} (X_\mathrm{A}) = x_{i}(X_\mathrm{A}) + u_{i} (X_\mathrm{A}),
\end{aligned}
\label{Eq:DeformDescription}
\end{equation}
where $u_{i}(X_\mathrm{A})$ is a small-amplitude incremental displacement associated with the deformation $B_\mathrm{e} \rightarrow B_\mathrm{t}$.


The deformation gradients arising from the deformations $B_\mathrm{0} \rightarrow B_\mathrm{t}$ and $B_\mathrm{0} \rightarrow B_\mathrm{e}$ are denoted by ${\boldsymbol{F}}$ and $\bar{\boldsymbol{F}}$ respectively and are defined by their components
\begin{equation}
\begin{aligned}
F_{i\mathrm{A}} = \frac{\partial \tilde{x}_{i}}{\partial X_\mathrm{A}}, \quad\quad \bar{F}_{i\mathrm{A}} = \frac{\partial x_{i}}{\partial X_\mathrm{A}}.
\end{aligned}
\label{Eq:DeformGradient}
\end{equation}
It then follows that
\begin{equation}
F_{i\mathrm{A}} = \bar{F}_{i\mathrm{A}}+u_{i,\mathrm{A}},
\label{2.3}
\end{equation}
where here and henceforth $u_{i,\mathrm{A}}=\paa u_i/\paa X_\mathrm{A}$.

\begin{figure}
\begin{center}
    \includegraphics[scale=0.28]{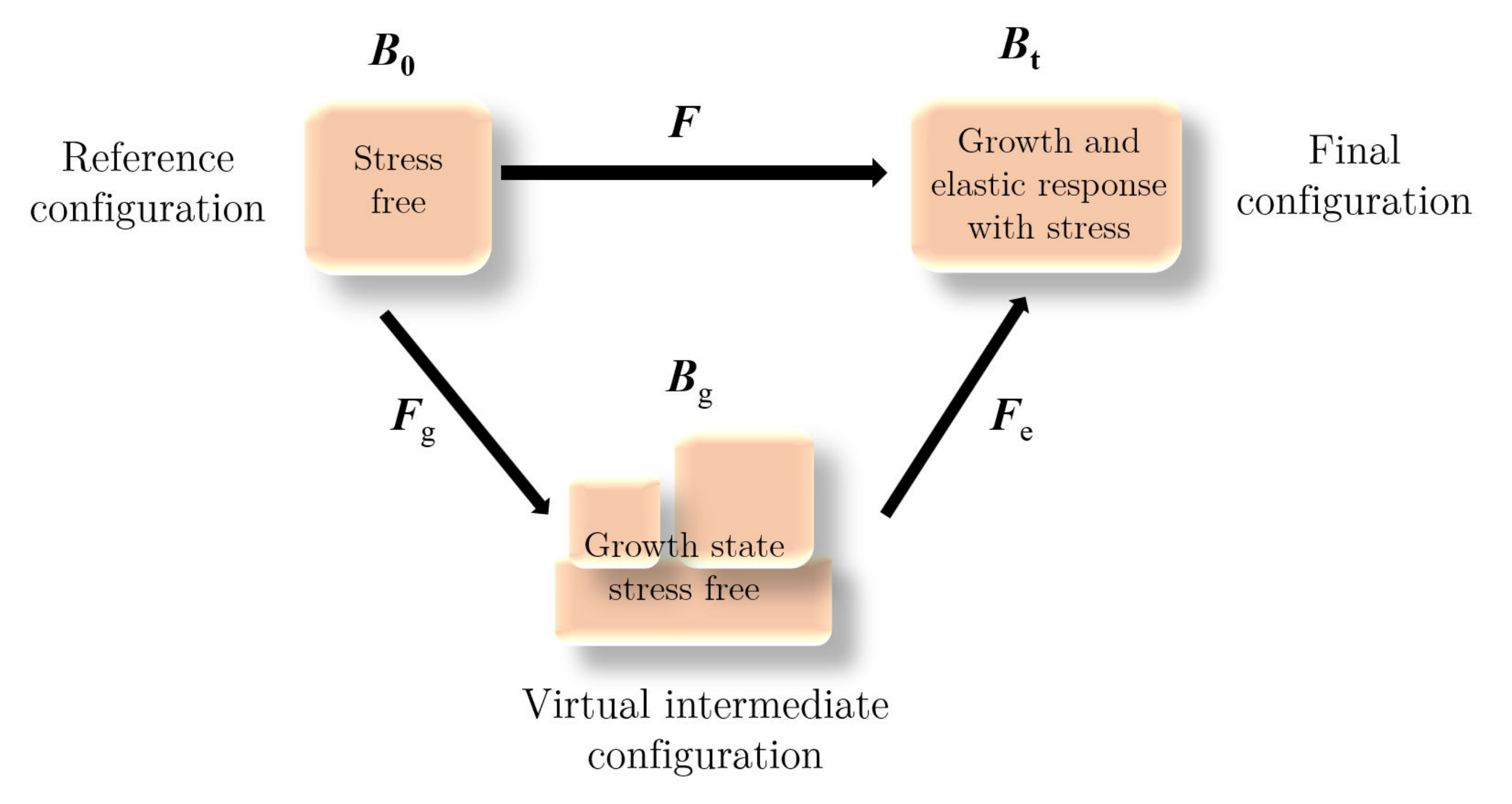}
    \caption{The multiplicative decomposition of the geometric deformation gradient tensor $\boldsymbol{F}$ into a growth part $\boldsymbol{F}_\mathrm{g}$  and an elastic part $\boldsymbol{F}_\mathrm{e}$. }
\label{fig:fig1}
\end{center}
\end{figure}

Following standard practice~\citep{amar2005growth,liu2014nonlinear,groh2022morphoelastic}, as shown in Figure~\ref{fig:fig1}, we assume that the deformation gradients in the growing elastic body are decomposed multiplicatively as
\begin{equation}
\begin{aligned}
\bar{\boldsymbol{F}}=\bar{\boldsymbol{F}}_\mathrm{e} \boldsymbol{F}_\mathrm{g}, \;\;\;\; \boldsymbol{F}=\boldsymbol{F}_\mathrm{e} \boldsymbol{F}_\mathrm{g},
\end{aligned}
\label{Eq:MultiPlicateGrow}
\end{equation}
where $\bar{\boldsymbol{F}_\mathrm{e}}$ and $\boldsymbol{F}_\mathrm{e}$ are the elastic deformation gradients and $\boldsymbol{F}_\mathrm{g}$ is the growth tensor.

Denote the strain energy {\it per unit volume in the grown state} by $W(\boldsymbol{F}_\mathrm{e})$. The Cauchy stress $\boldsymbol{\sigma}$ and nominal stress $\boldsymbol{S}$ are computed according to  $\boldsymbol{\sigma}=J^{-1} \boldsymbol{F}_\mathrm{e} \paa W/\paa \boldsymbol{F}_\mathrm{e}$ and $\boldsymbol{S}=({\rm det}\, \boldsymbol{F}) \boldsymbol{F}^{-1} \boldsymbol{\sigma}$, respectively, where $J={\rm det}\, \boldsymbol{F}_\mathrm{e}$. Thus, we have~\citep{amar2005growth}
\begin{equation}
\begin{aligned}
\boldsymbol{S}=({\rm det}\, \boldsymbol{F}_\mathrm{g}) \, \boldsymbol{F}_\mathrm{g}^{-1} \frac{\paa W(\boldsymbol{F}_\mathrm{e})}{\paa \boldsymbol{F}_\mathrm{e}}=\frac{\paa \bar{W}}{\paa \boldsymbol{F}},
\end{aligned}
\end{equation}
where $\bar{W}$ is known as the {\it augmented energy density function}~\citep{CiarlettaGrowth2012} and is defined by $\bar{W}(\boldsymbol{F}, \boldsymbol{F}_\mathrm{g})=({\rm det}\, \boldsymbol{F}_\mathrm{g}) W(\boldsymbol{F} \boldsymbol{F}_\mathrm{g}^{-1})$.

We assume that the strain energy function is given by
\begin{equation}
\begin{aligned}
W=\frac{\mu}{2} (I_1-2) -2 {\rm log}(J)+ \frac{\mu \nu}{1- 2 \nu} ({\rm log}(J))^2,
\end{aligned}
\label{Eq:ene1}
\end{equation}
where $I_1={\rm tr}\, (\boldsymbol{F}_\mathrm{e}^\top \boldsymbol{F}_\mathrm{e})={\rm tr}\,(\boldsymbol{F}^\top \boldsymbol{F} \boldsymbol{F}_\mathrm{g}^{-1} \boldsymbol{F}_\mathrm{g}^{-\top})$, and the $\mu$ and $\nu$ correspond to the ground-state shear modulus and Poisson's ration, respectively.

Throughout this paper, we consider isotropic growth so that $\boldsymbol{F}_\mathrm{g}=\mathrm{diag}(1+g, 1+g, 1)$ with $g$ denoting the scalar-valued growth factor and
\begin{equation}
J=(1+g)^{-2} {\rm det}\, \boldsymbol{F}, \;\;\;\; I_1=(1+g)^{-2} {\rm tr}\,(\boldsymbol{F}^\top \boldsymbol{F}).
\end{equation}
Expanding $\boldsymbol{S}$ around $\boldsymbol{F}=\bar{\boldsymbol{F}}$, we obtain
\begin{equation}
\dot{S}_{\mathrm{A}i}\equiv S_{\mathrm{A}i}- S^{(0)}_{\mathrm{A}i} =E^{(1)}_{i\mathrm{A}j\mathrm{B}} u_{j,\mathrm{B}}+\frac{1}{2} E^{(2)}_{i\mathrm{A}j\mathrm{B}k\mathrm{C}}u_{j,\mathrm{B}}u_{k,\mathrm{C}} + \frac{1}{6} E^{(3)}_{i\mathrm{A}j\mathrm{B}k\mathrm{C}m\mathrm{D}}u_{j,\mathrm{B}}u_{k,\mathrm{C}}u_{m,\mathrm{D}}+\cdots,
\label{const1}
\end{equation}
 where the first equation defines the stress increment $\dot{S}_{\mathrm{A}i}$ and
 \begin{equation}
 S^{(0)}_{\mathrm{A}i}=\left. \frac{\paa \bar{W}}{\paa F_{i\mathrm{A}}}\right|_{F=\bar{F}}, \;\;\;\;
 E^{(1)}_{i\mathrm{A}j\mathrm{B}}=\left.\frac{\paa^2 \bar{W}}{\paa F_{i\mathrm{A}} \paa F_{jB}}\right|_{F=\bar{F}}, \;\;\;\;
E^{(2)}_{i\mathrm{A}j\mathrm{B}k\mathrm{C}}=\left.\frac{\paa^3 \bar{W}}{\paa F_{i\mathrm{A}} \paa F_{j\mathrm{B}}\paa F_{k\mathrm{C}}}\right|_{F=\bar{F}}, \;\;{\rm etc.} \label{moduli}
\end{equation}
In the absence of body forces, the equations of equilibrium  are given by
\begin{equation}
\begin{aligned}
\dot{S}_{\mathrm{A}i,\mathrm{A}} = 0.
\end{aligned}
\label{Eq:EquilCondition} 
\end{equation}

We now specialise the above governing equations to the film/substrate bilayer system defined by $-\infty < X_2 <h_\mathrm{f}$ with $X_2=0$ corresponding to the interface and $h_\mathrm{f}$ being the film thickness. The (ground-state) shear modulus and growth factor for the film and substrate are denoted by $(\mu_\mathrm{f}, g_{\rm f})$ and $(\mu_\mathrm{s}, g_{\rm s})$, respectively. For simplicity, we assume that the Poisson's ratio of the film and substrate are identical. We define the modulus ratio $r_\mathrm{m}=\mu_\mathrm{s}/\mu_\mathrm{f}$ and the substrate/film growth ratio $r_\mathrm{g}=g_{\rm s}/g_{\rm f}$.
When we state an equation valid for both the film and substrate, we shall use symbols such as $\mu$ and $g$ without a subscript.

\subsection{Linear buckling analysis}
\label{Sec:AnalyticalLBA}
The linearised problem consists of solving the equilibrium equations
\begin{equation}
E^{(1)}_{iAjB} u_{j,AB} =0, \;\;\;\; -\infty < X_2 <h_\mathrm{f} \label{lead1}
\end{equation}
subject to the traction-free boundary conditions
\begin{equation}
E^{(1)}_{i2jB} u_{j,B}=0, \;\;\;\; X_2=h_\mathrm{f},
\label{tract-free} \end{equation}
the continuity conditions
\begin{equation}
\begin{aligned}
\left[ E^{(1)}_{i2jB} u_{j,B} \right]=0, \;\;\;\; [u_i]=0, \;\;\;\; X_2=0,
\end{aligned}
\label{Eq:cont1}
\end{equation}
and the decay conditions
\begin{equation}
u_i \to 0, \;\;\;\;{\rm as}\;\; X_2 \to -\infty.
\end{equation}
In Eq.~\eqref{Eq:cont1}, the square brackets denote the jump of the enclosed quantity as the interface is crossed.

We look for a normal mode solution of the form
\begin{equation}
{\bm u}={\bm v}(X_2) {\rm e}^{\ii k X_1}+ {\rm c.c},
\label{normal-mode}
\end{equation}
where $k$ is the wavenumber and c.c. denotes the complex conjugate of the preceding term. 
The reduced eigenvalue problem for the unknown vector function ${\bm v}(X_2)$ is solved with the aid of Mathematica. The resulting bifurcation condition is given by the determinant of a $6\times 6$ matrix equal to zero: 
\begin{equation}
\begin{aligned}
 f(g_{\rm f},  r_{\rm m}, r_{\rm g},  kh_\mathrm{f})=0.
\end{aligned}
\label{Eq:bif0}
\end{equation}
For each fixed modulus ratio $r_\mathrm{m}$ and growth ratio $r_\mathrm{g}$, this equation defines $g_{\rm f}$ implicitly as a function of $kh_\mathrm{f}$, typical graphs of which will be presented later. A local minimum may exist if $d g_{\rm f}/d (kh_\mathrm{f})=0$ for some non-zero $kh_\mathrm{f}$.
We denote by $(kh_\mathrm{f})_{\rm cr}$ the value of $kh_\mathrm{f}$ where $g_{\rm f}$ attains its global minimum
over the interval $kh_\mathrm{f} \in [0, \infty)$, and refer to the associated linear buckling solution as the critical mode. As will be seen later,  $g_{\rm f}$ may or may not have a local minimum and when a local minimum exists,
the global minimum may correspond to this local minimum or the value at $kh_\mathrm{f}=0$ (it can be shown that for the cases with $r_{\rm g}>1$, the global minimum cannot be attained in the limit $kh_\mathrm{f} \to \infty$).
%
The limit
$kh_\mathrm{f} \to 0$ can be achieved by $h_\mathrm{f} \to 0$, in which case the bilayer reduces to a homogeneous half-space and so it is appropriate to refer to the associated buckling mode as a Biot mode \citep{biot1963surface}. When $h_\mathrm{f}$ is fixed, as in our numerical simulations, the limit $kh_\mathrm{f} \to 0$ can only be achieved by $k \to 0$, and hence we refer to the associated mode more precisely as a {\it Biot mode with infinite wavelength}. Therefore, the
critical mode may be a sinusoidal mode with $(kh_\mathrm{f})_{\rm cr} \ne 0$ or a Biot mode with infinite wavelength.

%
%
%

We first present the asymptotic solutions of the exact bifurcation condition \rr{Eq:bif0} for both uniform and differential growth. The full solutions of this bifurcation condition will be discussed in the context of Figures \ref{fig:fig4}  and \ref{fig:CriticalModeChangeAnalytical} in Section 3 later.
\subsubsection{Uniform growth}
This corresponds to the case when the growth factors for the film and substrate are identical, i.e.\ $g_\mathrm{f}=g_\mathrm{s}=g$. The problem is then equivalent to the mechanical loading scenario with $\lambda=1/(1+g)$, where $\lambda$ is the stretch in the $X_1$-direction. To quantify this scenario, we define the nominal strain as $\varepsilon=1-\lambda$. We found that when $r_\mathrm{m}$ is small, the  strain $\varepsilon$ can be expanded as
\begin{equation}
\begin{aligned}
\varepsilon=&\frac{2 (\nu-1)^2}{3-4 \nu} \frac{r_\mathrm{m}}{k h_\mathrm{f}} +\frac{1}{12} (kh_\mathrm{f})^2+d_0 r_\mathrm{m}+d_1 (k h_\mathrm{f})^4+d_2 r_\mathrm{m} (k h_\mathrm{f}) +d_3 \frac{ r_\mathrm{m}^2}{ (kh_\mathrm{f})^2}\\
&+d_4 (kh_\mathrm{f})^2 r_\mathrm{m}+d_5 \frac{r_\mathrm{m}^2}{kh_\mathrm{f}}+O((kh_\mathrm{f})^6),
\end{aligned}
\label{Eq:UniGrowApp}
\end{equation}
where
\begin{equation}
\begin{aligned}
d_0&=\frac{(1-\nu ) (1-2 \nu)}{3-4\nu},\;\;\;\;
d_1=\frac{20 \nu ^3-29 \nu^2+28 \nu -9}{1440 (\nu -1)^2},
\;\;\;\;
d_2=\frac{35-120\nu+124 \nu^2-32  \nu^3}{24 (3-4 \nu )^2}, \\ 
d_3&=\frac{2 (\nu -1)^2 \left(11-32 \nu+31 \nu^2-16 \nu^4\right)}{(4 \nu-3)^3}, \;\;\;\;
d_4=\frac{1+6 \nu-12 \nu^2}{24 (4 \nu -3)^2}, \label{0.6} \\
d_5&=\frac{23-163 \nu+408 \nu^2-404 \nu^3+8 \nu^4+256 \nu^5-128 \nu^6}{2 (4 \nu -3)^3}.
\end{aligned}
\end{equation}

Substituting Eq.~\eqref{Eq:UniGrowApp} into the critical condition $\mathrm{d} \varepsilon/\mathrm{d} (kh_\mathrm{f})=0$ yields the critical wrinkling wavelength:
\begin{equation}
kh_\mathrm{f}=(kh_\mathrm{f})_{\rm cr} \equiv g_1 r_\mathrm{m}^{1/3}+ g_2 r_\mathrm{m} + g_3 r_\mathrm{m}^{4/3}+ O(r_\mathrm{m}^{5/3}),
\label{Eq:khUniformGrowthApp}
\end{equation}
with
\begin{equation}
\begin{aligned}
g_1&=2^{2/3} \sqrt[3]{3}\sqrt[3]{\frac{(\nu-1)^2}{3-4 \nu }}, \;\;\;\; g_2=\frac{-507+1400\nu-1064 \nu^2-544 \nu^3+960 \nu^4}{60 (3-4 \nu )^2}, \\
g_3&=\frac{(1-2 \nu)(-25+80 \nu-72 \nu^2-32 \nu^3+64 \nu^4)}{2\ 3^{2/3} (3-4\nu )^2 \sqrt[3]{8 \nu^2-14 \nu +6}}.
\end{aligned}
\label{Eq:app3}
\end{equation}
By substituting Eq.~\eqref{Eq:khUniformGrowthApp} back into Eq.~\eqref{Eq:UniGrowApp}, we can derive the expression for the critical strain:
\begin{equation}
\varepsilon_{\rm cr}=\frac{3^{2/3} (1-\nu)^{4/3}}{(6-8 \nu )^{2/3}} r_\mathrm{m}^{2/3}+\frac{(1-\nu) (2 \nu -1)}{4\nu -3} r_\mathrm{m} + g_4 r_\mathrm{m}^{4/3} +g_5 r_\mathrm{m}^{5/3}+O(r_\mathrm{m}^{6/3}),
\label{Eq:epsilonCrUniformGrowth}
\end{equation}
where
\begin{equation}
\begin{aligned}
g_4&= \frac{(1-\nu )^{2/3} \left(240 \nu ^4+16 \nu ^3-199 \nu^2+160 \nu -72\right)}{20\sqrt[3]{2}\cdot 3^{2/3} (3-4 \nu)^{7/3}}, \\
g_5&=\frac{128 \nu ^6-256 \nu ^5-20\nu ^4+434 \nu ^3-431 \nu^2+167 \nu -22}{2\cdot 2^{2/3}\sqrt[3]{3} (3-4 \nu )^{8/3}(1-\nu )^{2/3}}.
\end{aligned}
\end{equation}
Similar expressions for $(kh_\mathrm{f})_{\rm cr}$ and $\varepsilon_{\rm cr}$ have previously been derived by \cite{wlf2023} for a slightly different strain energy function.
In the limit of incompressibility ($\nu \to 0.5$), the expressions \eqref{Eq:epsilonCrUniformGrowth} and \eqref{Eq:khUniformGrowthApp} reduce to
\begin{equation}
\varepsilon_{\rm cr}=\frac{1}{4} (3r_\mathrm{m})^{2/3}-\frac{33}{160} r_\mathrm{m} (3r_\mathrm{m})^{1/3}+\frac{1}{24} (3r_\mathrm{m})^{5/3} +O(r_\mathrm{m}^2),
\label{Eq:UniGrowAppIncomp} 
\end{equation}
\begin{equation}
(kh_\mathrm{f})_{\rm cr}= (3r_\mathrm{m})^{1/3}+\frac{3}{20} r_\mathrm{m}+O(r_\mathrm{m}^{5/3}).
\label{app7}
\end{equation}
These two expressions without the $O(r_\mathrm{m}^{5/3})$ term in Eq.~\eqref{Eq:UniGrowAppIncomp} have previously been given by \cite{cai-fu1999}. Note that the extra $O(r_\mathrm{m}^{5/3})$ term in Eq.~\eqref{Eq:UniGrowAppIncomp} arises from the  inclusion of the $O((kh_\mathrm{f})^5)$ terms in Eq.~\eqref{Eq:UniGrowApp}. These additional higher-order terms improve the accuracy of $g_{\rm cr}$, but not the accuracy of $(kh_\mathrm{f})_{\rm cr}$ since $g_3 \to 0$ when $\nu \to 0.5$.






\subsubsection{Differential growth}
Following the procedure in the preceding section for uniform growth ($r_\mathrm{g}=1$),  we can determine the critical states (\emph{e.g.}\ critical growth factor in the film and the critical wavenumber) for each specified growth ratio $r_\mathrm{g}$:
\begin{equation}
\begin{aligned}
    g_{\rm f,cr}=\frac{3^{2/3} (1-\nu )^{4/3} }{(6-8 \nu)^{2/3}}r_\mathrm{m}^{2/3}+\frac{(1-\nu) (2 \nu -1)}{4\nu -3} r_\mathrm{m} + g_4^* r_\mathrm{m}^{4/3} +g_5^* r_\mathrm{m}^{5/3}+O(r_\mathrm{m}^{6/3}),
\end{aligned}
\label{Eq:GfcrDiffGrowth}
\end{equation}
\begin{equation}
\begin{aligned}
(kh_\mathrm{f})_{\rm cr}=g_1 r_\mathrm{m}^{1/3}+g_2^* r_\mathrm{m}+g_3^* r_\mathrm{m}^{4/3} +O(r_\mathrm{m}^{5/3}),
\end{aligned}
\label{Eq:khcrDiffGrowth}
\end{equation}
where $g_1$ is the same as in Eq.~\eqref{Eq:app3},
\begin{equation}
g_2^*=\frac{15 \left(16 \nu ^3-8 \nu ^2-1\right) r_g+4 (4 \nu -3) \left(60 \nu ^3-4 \nu ^2-62 \nu +41\right)}{60 (3-4 \nu )^2},
\label{Eq:DiffGrowthg2s}
\end{equation}

\begin{equation}
g_3^*=\frac{(1-2 \nu ) \left\{\left(16 \nu ^3-8 \nu ^2-1\right)
   r_g+8 \left(8 \nu ^4-6 \nu ^3-8 \nu ^2+10 \nu
   -3\right)\right\}}{2 \,\sqrt[3]{2}\, 3^{2/3} (3-4 \nu )^{7/3}
   \sqrt[3]{1-\nu }},
\label{Eq:DiffGrowthg3s}
\end{equation}
\begin{equation}
g_4^*=\frac{(1-\nu )^{2/3} \left\{15 \left(16 \nu ^3-8 \nu
   ^2-1\right) r_{\rm g}+(4 \nu -3) \left(60 \nu ^3-101 \nu ^2+152 \nu -71\right)\right\}}{20 \sqrt[3]{2}\, 3^{2/3}
   (3-4 \nu )^{7/3}},
\label{g4star}
\end{equation}
\begin{equation}
g_5^*=\frac{ \left(16 \nu ^4-88 \nu ^3+124
   \nu ^2-79 \nu +20\right) r_g+2 (1-2 \nu ) (4 \nu -3) \left(8 \nu ^3-3 \nu ^2-2 \nu
   +1\right)}{ 2^{5/3} \sqrt[3]{3} (1-\nu )^{-1/3}
   (3-4 \nu )^{8/3}}.
\label{g5star}
\end{equation}
Note that in the limit $r_\mathrm{g}\to 1$ (uniform growth), Eq.~\eqref{Eq:khcrDiffGrowth}  reduces to Eq.~\eqref{Eq:khUniformGrowthApp}.

It can be seen from Eqs.~\eqref{Eq:GfcrDiffGrowth} and~\eqref{Eq:khcrDiffGrowth} that the leading-order term in the critical wavenumber and the first two terms for the critical growth factor in the film are independent of the growth ratio $r_\mathrm{g}$. This is in accordance with the previous numerical results given by \citet{Dortdivanlioglu2017}, who found that the critical growth factor $g_\mathrm{f,cr}$ is almost independent of the substrate/film growth ratio $r_\mathrm{g}$ in the regime $0<r_\mathrm{g}<1$. In the double limits $r_\mathrm{g} \to 0$ (only film grows) and $\nu \to 0.5$ (incompressible material), the above expressions are consistent with Eqs.~(3.11) and (3.12) in \cite{alawiye2019revisiting}.

\subsection{Weakly nonlinear analysis for the case when $(kh_\mathrm{f})_{\rm cr}\ne 0$}
As remarked earlier, the critical mode may be a Biot mode with $(kh_\mathrm{f})_{\rm cr}=0$ or a sinusodal mode with $(kh_\mathrm{f})_{\rm cr} \ne 0$. Drawing upon what is already known for the buckling of a single half-space (the Biot problem) and a film/substrate bilayer under mechanical compression \citep{fwf2023}, we expect that the bifurcation associated with a Biot mode in the current context is also subcritical and will lead to a creased configuration. Thus, our weakly nonlinear analysis will be focused on the cases with $(kh_\mathrm{f})_{\rm cr}\ne 0$.

When $h_\mathrm{f}$ is specified, the wavenumber of the critical mode, denoted by  $k_{\rm cr}$, is computed from
$k_{\rm cr}=(kh_\mathrm{f})_{\rm cr}/h_\mathrm{f}$. We then need to determine whether the associated bifurcation is subcritical or supercritical through a weakly nonliner analysis. To this end, we denote the critical values of $g_{\rm f}$ and $g_{\rm s}$ by $g_{{\rm f,cr}}$ and $g_{{\rm s,cr}}$, respectively, and let
\begin{equation}
\begin{aligned}
g_{\rm f}=g_{{\rm f,cr}}+\ep^2 \hat{g}_{\rm f}, \;\;\;\; g_{\rm s}=g_{{\rm s,cr}}+\ep^2 \hat{g}_{\rm s},
\end{aligned}
\label{Eq:gg1}
\end{equation}
where $\hat{g}_{\rm f}$ and $\hat{g}_{\rm s}$ are constants and $\ep$ is a small parameter characterising the deviation of $g_{\rm f}$ and $g_{\rm s}$ from their critical values. In the case when the ratio $r_{\rm g}=g_{\rm s}/g_{\rm f}$ is fixed, $g_{\rm f}$ is viewed as the
control parameter and then the expression for $g_{\rm s}$ is not independent. We look for an asymptotic solution of the form
\begin{equation}
\begin{aligned}
\tilde{\bm x}=\left\{ \begin{array}{c}  X_1 \\ \lambda_2 X_2 \end{array}\right\}+{\bm u}, \;\;\;\; {\bm u}=\ep {\bm u}^{(1)}+\ep^2 {\bm u}^{(2)}+\ep^3 {\bm u}^{(3)}+\cdots,
\end{aligned}
\label{Eq:asy1}
\end{equation}
where the leading order solution is given by
\begin{equation}
{\bm u}^{(1)}= A {\bm v}(X_2) {\rm e}^{\ii k_{\rm cr} X_1}+ {\rm c.c}, \label{leading-order} \end{equation}
and the other terms ${\bm u}^{(2)}, {\bm u}^{(2)}, ...$ are functions of $X_1$ and $X_2$ to be determined at successive orders. Note that
equation \rr{leading-order} is simply the linear solution evaluated at the critical state, with $A$ denoting the undetermined amplitude.
The stretch $\lambda_2$ in Eq.~\eqref{Eq:asy1}$_1$ induced by growth is dependent on $g_{\rm f}$ and $g_{\rm s}$ and so has a similar expansion to Eq.~\eqref{Eq:gg1}. 

On substituting Eqs.~\eqref{Eq:gg1} and \eqref{Eq:asy1} into Eq.~\eqref{const1}, we obtain
\begin{equation}
\begin{aligned}
S_{Ai}=&\bar{S}^{(0)}_{Ai}+\ep \bar{E}^{(1)}_{iAjB} u^{(1)}_{j,B}+\ep^2 \left\{ \hat{g} \bar{S}^{(0)'}_{Ai}   + \bar{E}^{(1)}_{iAjB} u^{(2)}_{j,B}  + \frac{1}{2} E^{(2)}_{iAjBkC}u^{(1)}_{j,B} u^{(1)}_{k,C} \right\} \\
&+\ep^3 \left\{ \bar{E}^{(1)}_{iAjB} u^{(3)}_{j,B}+ \hat{g} \bar{E}^{(1)'}_{iAjB} u^{(1)}_{j,B} +  E^{(2)}_{iAjBkC}u^{(1)}_{j,B} u^{(2)}_{k,C}+    \frac{1}{6} E^{(3)}_{iAjBkCmD} u^{(1)}_{j,B}u^{(1)}_{k,C}u^{(1)}_{m,D} \right\}+\cdots,
\end{aligned}
\label{const11}
\end{equation}
where
$$
\bar{S}^{(0)}_{Ai}= S^{(0)}_{iA}|_{g=g_{\rm cr}}, \;\;\;\; \bar{S}^{(0)'}_{Ai}= \left.  \frac{\mathrm{d} \bar{S}_{Ai}}{\mathrm{d} g}\right|_{g=g_{\rm cr}},  $$
$$
\bar{E}^{(1)}_{iAjB}= ({E}^{(1)}_{iAjB})|_{g=g_{\rm cr}}, \;\;\;\; \bar{E}^{(1)'}_{iAjB}= \left.  \frac{\mathrm{d} {E}^{(1)}_{iAjB} }{\mathrm{d} g}\right|_{g=g_{\rm cr}}.  $$
The above expressions are valid for both the film and substrate. When applying them to the film, for instance, we simply replace $g_{\rm cr}$ by $g_{\rm f,cr}$ and $\hat{g}$ by $\hat{g}_{\rm f}$.

The leading-order problem is automatically satisfied since it is the same problem as in the linear buckling analysis. At the second order, we need to solve
\begin{equation}
  \bar{E}^{(1)}_{iAjB} u^{(2)}_{j,AB} =\hat{R}_i, \;\;\;\; -\infty < X_2 <h_\mathrm{f} \label{lead1f} \end{equation}
subject to the traction-free boundary conditions
\begin{equation}
\bar{E}^{(1)}_{i2jB} u^{(2)}_{j,B}=\hat{S}_i, \;\;\;\; X_2=h_\mathrm{f},
\label{tract-free-1}
\end{equation}
the continuity conditions
\begin{equation}
[  \bar{E}^{(1)}_{i2jB} u^{(2)}_{j,B} ]=\hat{G}_i, \;\;\;\; [u^{(2)}_i]=0, \;\;\;\; X_2=0,
\label{cont11f}
\end{equation}
and the decay conditions
\begin{equation}
u^{(2)}_i \to 0, \;\;\;\;{\rm as}\;\; X_2 \to -\infty,
\label{decay1}
\end{equation}
where the inhomogeneous terms $\hat{R}_i, \hat{S}_i, \hat{G}_i$ only involve the leading-order solution.
This boundary value problem can be solved without having to impose any solvability conditions.

At third order, the boundary value problem consists of solving
\begin{equation}
\bar{E}^{(1)}_{iAjB} u^{(3)}_{j,AB} ={R}_i, \;\;\;\; -\infty < X_2 <h_\mathrm{f}
\label{lead1g}
\end{equation}
subject to the traction-free boundary conditions
\begin{equation}
\bar{E}^{(1)}_{i2jB} u^{(3)}_{j,B}={S}_i, \;\;\;\; X_2=h_\mathrm{f},
\label{tract-free-1f}
\end{equation}
the continuity conditions
\begin{equation}
[  \bar{E}^{(1)}_{i2jB} u^{(3)}_{j,B} ]={G}_i, \;\;\;\; [u^{(3)}_i]=0, \;\;\;\; X_2=0,
\label{cont11}
\end{equation}
and the decay conditions
\begin{equation}
u^{(3)}_i \to 0, \;\;\;\;{\rm as}\;\; X_2 \to -\infty,
\label{decay2}
\end{equation}
where the right hand sides $R_i$, $S_i$ and $G_i$ are dependent on the leading and second order solutions.
A solvability condition must be imposed at this order due to self resonance. This condition yields an amplitude equation that is derived using the virtual work method \citep{fu1995} as follows.



We first denote the domains occupied by the film and substrate over one period $2\pi/k_\mathrm{cr}$ by $B_{\rm f}$ and $B_{\rm s}$, respectively. On the one hand, we have
\begin{equation}
\begin{aligned}
&\int_{B_{\rm f} \cup B_{\rm s}}   \bar{E}^{(1)}_{iAjB} u^{(3)}_{j,B} u^{(0)}_{i,A} \mathrm{d}X_1\mathrm{d}X_2 \\
&=\int_{B_{\rm f} \cup B_{\rm s}} \left(  \bar{E}^{(1)}_{iAjB}  u^{(3)}_{j} u^{(0)}_{i,A}\right)_{,B} \mathrm{d}X_1\mathrm{d}X_2 -\int_{B_{\rm f} \cup B_{\rm s}}   \bar{E}^{(1)}_{iAjB} u^{(3)}_{j} u^{(0)}_{i,AB} \mathrm{d}X_1\mathrm{d}X_2  \\
&= \int_{\paa B_{\rm f} \cup \paa B_{\rm s}}    \bar{E}^{(1)}_{iAjB} u^{(0)}_{i,A}  u^{(3)}_{j}  N_B \mathrm{d}S -\int_{B_{\rm f} \cup B_{\rm s}}   \bar{E}^{(1)}_{jBiA} u^{(0)}_{i,AB}  u^{(3)}_{j}  \mathrm{d}X_1\mathrm{d}X_2 =0,
\end{aligned}
\end{equation}
where we have used the fact that $u^{(0)}$ is the leading-order solution so that it satisfies the equilibrium equation $  \bar{E}^{(1)}_{iAjB}u^{(0)}_{i,AB}=0$ and the auxiliary conditions.

On the other hand, writing $\pi^{(3)}_{iA}=  \bar{E}^{(1)}_{iAjB} u^{(3)}_{j,B}$, we may evaluate the same integral differently as follows:
\begin{equation}
\begin{aligned}
  \int_{B_{\rm f} \cup B_{\rm s}}   \bar{E}^{(1)}_{iAjB} u^{(3)}_{j,B} u^{(0)}_{i,A} \mathrm{d}X_1\mathrm{d}X_2 = &\int_{\paa B_{\rm f} \cup \paa B_{\rm s}} \pi^{(3)}_{iA} u^{(0)}_{i}N_A \mathrm{d}S-\int_{B_{\rm f} \cup B_{\rm s}} \pi^{(3)}_{iA,A} u^{(0)}_{i} \mathrm{d}X_1\mathrm{d}X_2\\
   = & \int_0^{2 \pi/k_\mathrm{cr}} \left.\pi^{(3)}_{i2}  u^{(0)}_{i}\right|_{x_2=h_\mathrm{f}} \mathrm{d}X_1 -  \int_0^{2 \pi/k_\mathrm{cr}} \left.\pi^{(3)}_{i2}  u^{(0)}_{i}\right|_{x_2=0^+} \mathrm{d}X_1 \\
  &+  \int_0^{2 \pi/k_\mathrm{cr}}\left. \pi^{(3)}_{i2}  u^{(0)}_{i}\right|_{x_2=0^-} \mathrm{d}X_1   - \int_{B_{\rm f} \cup B_{\rm s}} R_i u^{(0)}_{i} \mathrm{d}X_1\mathrm{d}X_2.
\end{aligned}
\end{equation}

Each integrand in the above expression consists of terms independent of $X_1$ and other terms proportional to ${\rm e}^{\ii X_1}$, ${\rm e}^{\ii 3 X_1}$ and their complex conjugates. Only terms independent of $X_1$ can survive the integration from $X_1=0$ to $X_1=2 \pi/k_\mathrm{cr}$. It then follows that
\begin{equation}
\left.\pi^{(3)}_{i2}  u^{(0)}_{i}\right|_{X_2=h_\mathrm{f}}   -    \left.\pi^{(3)}_{i2}  u^{(0)}_{i}\right|_{X_2=0^+}  +  \left. \pi^{(3)}_{i2}  u^{(0)}_{i}\right|_{X_2=0^-}- \int^{0}_{-\infty}R_i u^{(0)}_{i} \mathrm{d}X_2 - \int^{H}_{0}R_i u^{(0)}_{i} \mathrm{d}X_2 =0,
\end{equation}
where it is understood that all terms that vanish after the above integration are dropped. Rearranging with the use of \eqref{tract-free-1f}, we obtain
\begin{equation}
\int^{0}_{-\infty}R_i u^{(0)}_{i} \mathrm{d}X_2 + \int^{H}_{0}R_i u^{(0)}_{i} \mathrm{d}X_2 + \left.S_i  u^{(0)}_{i}\right|_{X_2=0^+} - \left.S_i  u^{(0)}_{i}\right|_{X_2=h_\mathrm{f}}  -  \left. S_i  u^{(0)}_{i}\right|_{X_2=0^-}=0,
\end{equation}
or equivalently,
\begin{equation}
\int^{0}_{-\infty}R_i u^{(0)}_{i} \mathrm{d} X_2 + \int^{H}_{0}R_i u^{(0)}_{i} \mathrm{d} X_2 + \left.[S_i]  u^{(0)}_{i}\right|_{X_2=0^+} - \left.S_i  u^{(0)}_{i}\right|_{X_2=h_\mathrm{h}}=0,
\end{equation}
where $R_i$ is defined by Eq.~\eqref{tract-free-1f}. With the further use of  Eq.~\eqref{cont11}, we finally arrive at
\begin{equation}
\int^{0}_{-\infty}R_i u^{(0)}_{i} \mathrm{d} X_2 + \int^{H}_{0}R_i u^{(0)}_{i} \mathrm{d}X_2 + \left.G_i  u^{(0)}_{i}\right|_{X_2=0^+} - \left.S_i  u^{(0)}_{i}\right|_{X_2=h_\mathrm{f}}=0.
\end{equation}
This equation can ultimately be manipulated into the form
\begin{equation}  \hat{g}_{\rm f} A-c_1 |A|^2 A=0, \label{amplitude} \end{equation}
where $c_1$ is a constant whose expression is not written out here for the sake of brevity. The non-trivial solution
$|A|=\sqrt{ \hat{g}_{\rm f}/c_1}$ can only exist for $\hat{g}_{\rm f}/c_1>0$. It thus follows that the bifurcation is supercritical if $c_1$ is positive and subcritical if $c_1$ is negative.

A Mathematica code is written to carry out the above calculations. For the case of uniform growth and both the film and substrate being neo-Hookean, the code reproduces the previously known result~\citep{cai-fu1999, hutchinson2013role, Alawiye2020} that bifurcation changes character (supercritical or subcritical) at $r_\mathrm{m}=0.57045$. For the more general cases, analogous results will be presented in the following discussions.

\subsection{Finite element modelling and bifurcation analysis}
\begin{figure}
\centering
\includegraphics[trim=0.0in 0.0in 0.0in 0.00in, clip=true, scale=1]{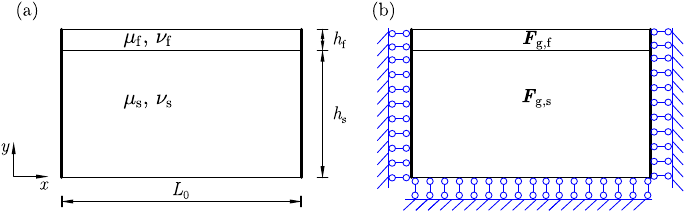}
\caption{(a) Geometry of the film/substrate growing bilayer for the FE model. (b) Boundary conditions of the growing bilayer in the FE model. $\boldsymbol{F}_\mathrm{g,f}$ and $\boldsymbol{F}_\mathrm{g,s}$ are the deformation gradient in the film and substrate due to growth, respectively (see also Figure~\ref{fig:fig1}). }
\label{fig:fig2}
\end{figure}

To verify the analytical model and further characterise the post-buckling behaviour in the fully nonlinear regime, we developed a nonlinear finite element (FE) model for growing bilayers using an in-house implementation~\citep{groh2018generalised,groh2022morphoelastic}. Sixteen-noded, isoparametric 2D plane strain elements with the growth variable applied at integration points are adopted to discretise the bilayer. The applied discretisation mesh is most dense towards the interface of the film/substrate and gradually coarsens towards the bottom of the substrate. The applied boundary condition is shown in Figure~\ref{fig:fig2}(b). Horizontal displacement restraints are applied at the left and right ends, and vertical displacement restraints are applied on the bottom of the substrate.

Note that in the developed FE model, the assumption of an infinitely deep substrate, as in the analytical model, cannot be implemented. To ensure that the FE model approximately satisfies this assumption and thus to minimise the influence of the vertical restraint on the substrate's bottom edge, a sensitivity study was conducted. The results indicate that when the critical wrinkling mode is sinusoidal, a substrate depth equal to thirty times the critical wrinkling wavelength yields accurate results (compared to the analytical results) with reasonable computational effort. However, when the critical mode is a Biot mode~\citep{biot1963surface} with an infinite characteristic wavelength, approximating the accurate solution requires a large (deep and wide) model. In fact, while the FE model does provide accurate predictions for the Biot wrinkling strain, it can never `truly' represent the Biot mode as the finite size of the FE model has embedded a characteristic length scale that is absent in the analytical model based on a semi-infinite halfspace. Given these circumstances, the FE model is mainly used to benchmark the analytical results and to explore the deep post-critical regime where asymptotic expansions become cumbersome.

The width of the FE model is set to be four times the critical wrinkling wavelength when the critical mode is sinusoidal. As far as the authors are aware, no generalised closed-form solutions exist to determine the critical strain $\varepsilon_\mathrm{cr}$ and critical wrinkling wavelength $L_\mathrm{crw}$ of growing bilayers with different growth rates in the film and the substrate. Hence, in the finite-width FE model, the critical wrinkling states, which correspond to sinusoidal wrinkling of a bilayer of infinite width, is determined using a minimisation process. The critical point for some assumed model width $L^*_0$ is determined by continuation and bifurcation point isolation, and the locus of this critical point is then traced by smoothly varying the model width until the value of $L_0$ is encountered for which $\varepsilon_\mathrm{cr}$ is minimised~\citep{Shen2022ProgramBilayer}. At this point the model width $L_0$ is an integer multiple of the critical wrinkling wavelength $L_\mathrm{crw}$. For all further simulations we then fix the width of the FE model to $L_0 = 4L_\mathrm{crw}$.


This procedure highlights some of the unique properties of the in-house FE code in that it provides the functionality of directly pinpointing bifurcation points and tracing these points through parameter space (\emph{e.g.}\ geometry or growth ratio) instead of relying on computationally expensive parametric studies~\citep{groh2018generalised,groh2022morphoelastic}. As a result, we can accurately determine the critical state and trace the post-critical behavior of growing bilayers without prior knowledge of possible branching events. 

\section{Critical states}
\label{Sec:CriticalState}
To compare our results with solutions on the wrinkling of bilayers under uniform mechanical compression, we show results using the nominal strain, $\varepsilon = g/(1+g)$, rather than the growth factor, $g$.
As mentioned in the introduction, we focus on bilayers with stiffness ratio of $\mu_\mathrm{f}/\mu_\mathrm{s}<50$.

\begin{table}[h]
\centering
\caption{Material and geometric properties of the growing film/substrate bilayer, where $L_\mathrm{crw}$ is the critical wrinkling wavelength; $g_\mathrm{s}$ and $g_\mathrm{f}$ are the isotropic growth factors in the substrate and film, respectively. }
\label{tab:bilayerSystemGeoMat}
\begin{tabular}{cccccccc}
\hline
$\mu_\mathrm{f}$ ($\mathrm{MPa}$)& $\mu_\mathrm{s}$ ($\mathrm{MPa}$) & $\nu_\mathrm{f}$ & $\nu_\mathrm{s}$ & $h_\mathrm{f}$ (mm) & $h_\mathrm{s}$&$L_0$ &$g_\mathrm{s}/g_\mathrm{f}$ \\ \hline
$0.0105-0.5$  & 0.01 & 0.49 & 0.49 & 0.4  & 30$L_\mathrm{crw}$&4$L_\mathrm{crw}$&$0-50$ \\ \hline
\end{tabular}
\end{table}

\subsection{Uniform growth}
\begin{figure}[]
\centering
\includegraphics[trim=0.0in 0.0in 0.0in 0.00in, clip=true, width=1.0\textwidth]{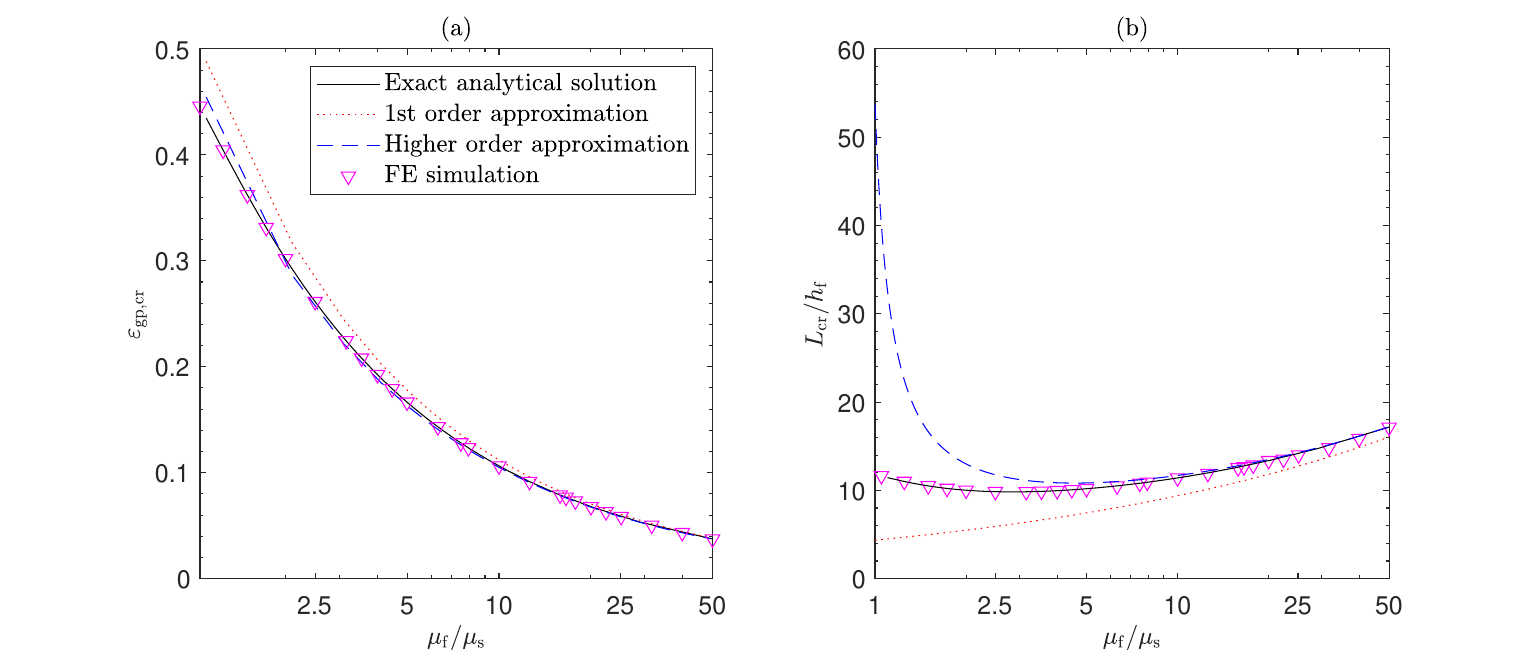}
\caption{(a) The nominal critical strain and (b) critical wavelength of bilayers under isotropic and uniform growth, i.e.\ $g_\mathrm{s}/g_\mathrm{f}=1$. Solid curves represent the solution of the exact bifurcation condition Eq.\eqref{Eq:bif0}; dashed and dotted curves represent the asymptotic solutions with first-order and higher-order approximation, respectively. Triangular symbols represent the solutions from the FE simulation. The geometry and material properties of the bilayers are shown in Table~\ref{tab:bilayerSystemGeoMat}. }
\label{fig:fig3}
\end{figure}

Figure~\ref{fig:fig3} presents the nominal critical strain and critical wavelength of bilayers with uniform growth ($g_\mathrm{s}/g_\mathrm{f}=1$) with the film/substrate modulus ratio ranging from 1 to 50. In general, the nominal critical strain decreases with increasing $\mu_\mathrm{f}/\mu_\mathrm{s}$. When the stiffness of the film and substrate is equal, i.e.\ $\mu_\mathrm{f}/\mu_\mathrm{s}=1$, the nominal critical strain is 0.458, i.e.\ the critical strain of the Biot mode \citep{biot1963surface}. The critical wrinkling wavelength first decreases with decreasing $\mu_\mathrm{f}/\mu_\mathrm{s}$, then reaches a minimum value at $\mu_\mathrm{f}/\mu_\mathrm{s}=2.5$ and finally increases again with further decreasing $\mu_\mathrm{f}/\mu_\mathrm{s}$. Theoretically, the critical wrinkling wavelength tends to infinity as $\mu_\mathrm{f}/\mu_\mathrm{s}\rightarrow1$~\citep{biot1963surface}. Note that the exact analytical solutions, which are solved numerically, show excellent correlation with the nonlinear FE simulation.

Apart from the exact analytical solution of the critical strain and critical wavelength solved numerically, we previously derived asymptotic solutions of different order. The results for the critical strain derived from Eq.~\eqref{Eq:epsilonCrUniformGrowth} and the critical wavelength derived from Eq.~\eqref{Eq:khUniformGrowthApp} are presented in Figures~\ref{fig:fig3}(a,b), respectively. The first-order asymptotic solutions of critical wavelength and critical strain, which have been reported to accurately describe the critical wrinkling state when $\mu_\mathrm{f}/\mu_\mathrm{s} \gg 1$~\citep{Cao2012,cheng2021intricate}, are no longer valid in the regime where $\mu_\mathrm{f}/\mu_\mathrm{s}<50$, as shown in Figure~\ref{fig:fig3}. The higher-order approximation shows better accuracy than the first-order approximation in the regime $5<\mu_\mathrm{f}/\mu_\mathrm{s}<50$ but the discrepancy with the exact analytical solutions increases when $\mu_\mathrm{f}/\mu_\mathrm{s}<5$, especially for the critical wavelength. The higher-order approximation captures the trend of increasing critical wavelength for $\mu_\mathrm{f}/\mu_\mathrm{s}<2.5$ but the accuracy is not satisfactory. Indeed it is this mode change for small values of $\mu_\mathrm{f}/\mu_\mathrm{s}$ that the asymptotic expansion based on an assumption of large $\mu_\mathrm{f}/\mu_\mathrm{s}$ inherently has difficulty capturing.

Considering the subcritical post-buckling behaviour of bilayers with film and substrate exhibiting comparable stiffness, creasing is likely to occur before the formation of stable wrinkling patterns. In other words, it is unlikely that sinusoidal wrinkling patterns will be observed in experiments in such arrangements.



\subsection{Differential growth}
In the preceding section, we studied the effect of changing $\mu_\mathrm{f}/\mu_\mathrm{s}$ on the critical wrinkling behaviour of bilayers under uniform film/substrate growth. We now investigate the effects of differential growth on the critical wrinkling wavelength and critical growth factor. Both film and substrate grow isotropically but at different rates ($g_\mathrm{f}$ and $g_\mathrm{s}$, respectively). Unlike prior work \citep{jin2019post,Alawiye2020}, where the substrate grows slower than the film ($r_\mathrm{g} = g_\mathrm{s}/g_\mathrm{f} \leq 1$), here we expand the parameter regime and put more focus on cases where the substrate grows faster than the film. In such cases, nonlinear effects from the substrate introduce interesting mechanics that have not been explored before.

\begin{figure}
\centering
\includegraphics[trim=0.0in 0.00in 0.0in 0.00in, clip=true, width=1.0\textwidth]{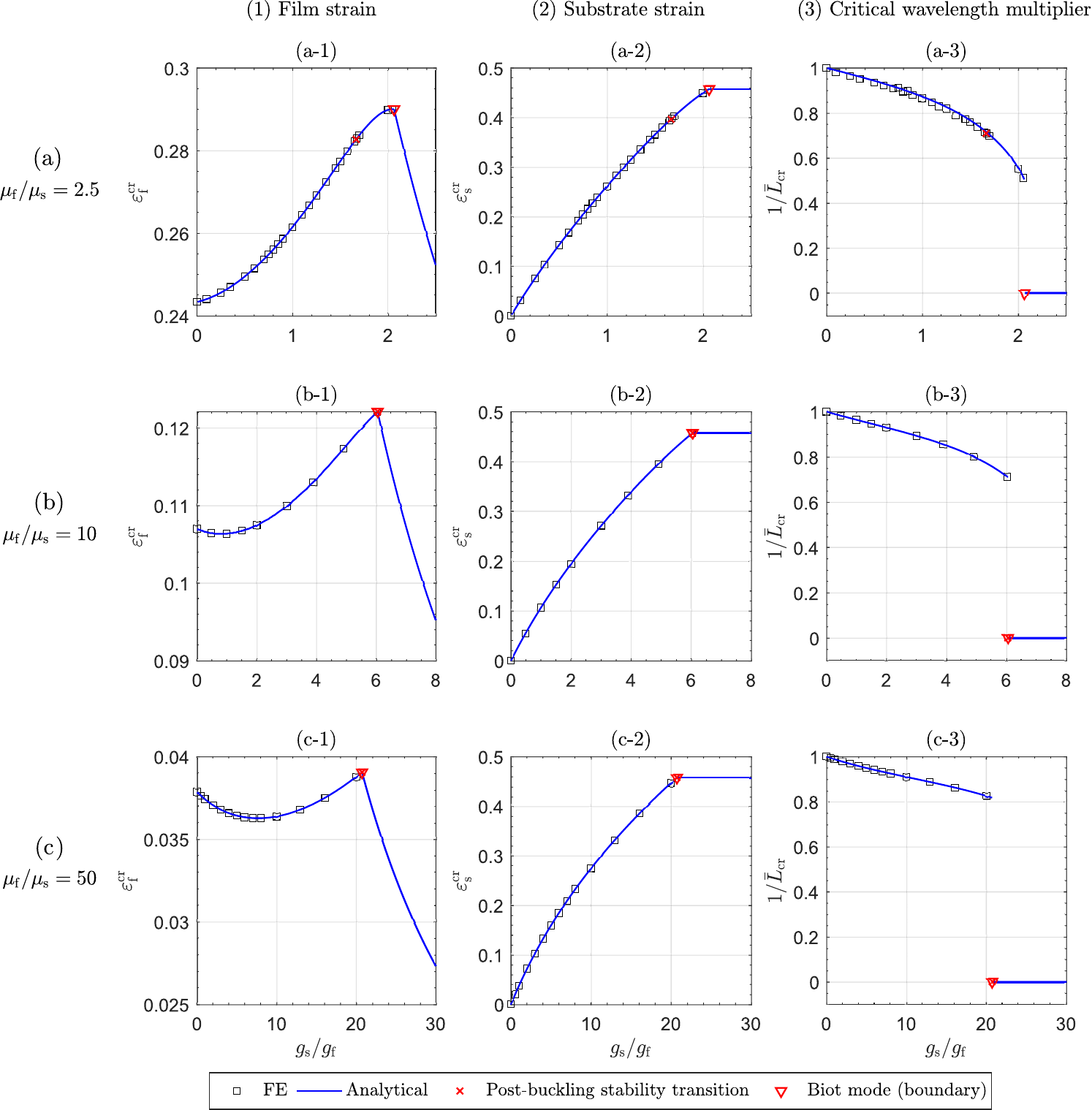}
\caption{The nominal critical strain and critical wavelength for bilayers with (a) $\mu_\mathrm{f}/\mu_\mathrm{s}=2.5$, (b) $\mu_\mathrm{f}/\mu_\mathrm{s}=10$, and (c) $\mu_\mathrm{f}/\mu_\mathrm{s}=50$ with varying substrate/film growth ratio $r_\mathrm{g}=g_\mathrm{s}/g_\mathrm{f}$. In each case, columns 1 and 2 are the nominal critical strain in the film and substrate, respectively, and column 3 is the reciprocal of the relative wavelength multiplier, with $\bar{L}_\mathrm{cr}=L_\mathrm{crw}/L_\mathrm{crw,0}$ and $L_\mathrm{cr,0}$ being the critical wrinkling wavelength at $g_\mathrm{s}/g_\mathrm{f}=0$, i.e.\ pure film growth. $\varepsilon^\mathrm{cr}_\mathrm{f}$ and $\varepsilon^\mathrm{cr}_\mathrm{s}$ are the nominal critical strains in the film and substrate, respectively. Solid curves represent the solutions of the exact bifurcation condition Eq.\eqref{Eq:bif0} and square symbols represent the solutions from FE simulations. Note that the FE model cannot model the case of an infinite half-space as considered in the analytical model. Hence, only the FE results before the Biot mode (red triangles) becomes the critical mode are presented. }
\label{fig:fig4}
\end{figure}

Figure~\ref{fig:fig4} presents the nominal critical strain in the film, substrate and critical wrinkling wavelength of bilayers with $\mu_\mathrm{f}/\mu_\mathrm{s}=2.5, 10, 50$ as a function of the growth ratio $r_\mathrm{g} = g_\mathrm{s}/g_\mathrm{f}$.
For all three values of $\mu_\mathrm{f}/\mu_\mathrm{s}$  the nominal critical strain in the film shows a sigmoidal relationship with $r_\mathrm{g}$, whereas the nominal critical strain in the substrate has a monotonic relationship with $r_\mathrm{g}$ and reaches a plateau for high $r_\mathrm{g}$. Similarly, the critical wrinkling wavelength monotonically increases with increasing $r_\mathrm{g}$ and becomes infinite above a certain threshold of $r_\mathrm{g}$. The peak value of $\varepsilon_\mathrm{f}^\mathrm{cr}$ in Figure~\ref{fig:fig4}(a), the start of the plateau in $\varepsilon_\mathrm{s}^\mathrm{cr}$ in Figure~\ref{fig:fig4}(b) and the sharp change from finite to infinite critical wavelength in Figure~\ref{fig:fig4}(c) all occur at the same value of $r_\mathrm{g}$ (for identical $\mu_\mathrm{f}/\mu_\mathrm{s}$) and correspond to the transition from sinusoidal wrinkling to Biot wrinkling. We define the corresponding $r_\mathrm{g}$ as \emph{the Biot threshold growth ratio} and with decreasing $\mu_\mathrm{f}/\mu_\mathrm{s}$, the Biot threshold growth ratio decreases, i.e.\ less differential growth is required to cause Biot wrinkling. Note that for cases with $\mu_\mathrm{f}/\mu_\mathrm{s}=10$ and $50$, $\varepsilon_\mathrm{f}^\mathrm{cr}$ initially decreases and then increases with increasing $g_\mathrm{s}/g_\mathrm{f}$. This nonlinear relationship between $r_\mathrm{g}$ and $\varepsilon^\mathrm{cr}$ precludes the use of a closed-form modification factor based on an effective film/substrate stiffness ratio, as done for a pre-stretched substrate~\citep{hutchinson2013role}.

\begin{figure}
\centering
\includegraphics[trim=1.0in 0.0in 1.0in 0.00in, clip=true, width=1.0\textwidth]{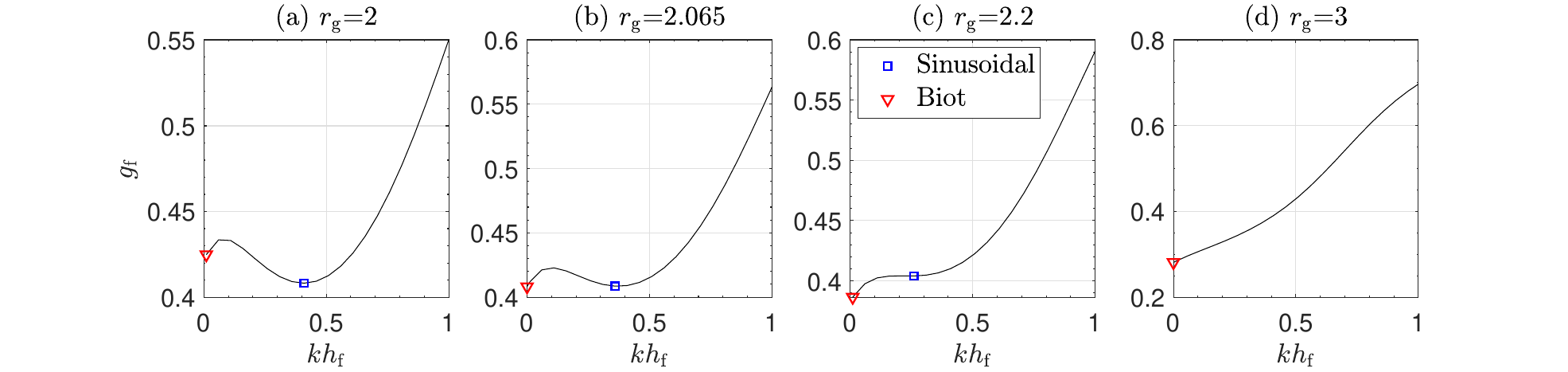}
\caption{Evolution of the critical growth factor and critical wrinkling wavelength for values of $r_\mathrm{g}(=g_\mathrm{s}/g_\mathrm{f}$) below, equal to, marginally above, and far above the Biot threshold growth ratio, computed with the use of the exact bifurcation condition Eq.~\eqref{Eq:bif0}. Note that $\mu_\mathrm{f}/\mu_\mathrm{s}=2.5$ for all cases, and the other geometry and material properties are as presented in Table~\ref{tab:bilayerSystemGeoMat}. }
\label{fig:CriticalModeChangeAnalytical}
\end{figure}

To shed light on the results in Figure~\ref{fig:fig4}, Figure~\ref{fig:CriticalModeChangeAnalytical} presents the evolution of the critical solution (growth factor versus  wrinkling wavelength) for different $r_\mathrm{g} = g_\mathrm{s}/g_\mathrm{f}$ from the analytical model for bilayers with $\mu_\mathrm{f}/\mu_\mathrm{s}=2.5$.
When $r_\mathrm{g}=2$, the local minimum at $kh_\mathrm{f}\neq0$ is the global minimum and thus the critical wrinkling wavelength is finite, i.e.\ we have a sinusoidal wrinkling mode.
When $r_\mathrm{g}=2.065$, the critical strain at the local minimum of $kh_\mathrm{f}\neq0$ is equal to the critical strain for $kh_\mathrm{f}=0$ (the Biot mode).
When $r_\mathrm{g}=2.2$, the global minimum switches to $kh_\mathrm{f}=0$ and the critical wrinkling wavelength is infinite, i.e.\ the Biot mode. While we retain the local minimum for $kh_\mathrm{f}\neq0$, this now corresponds to the second bifurcation point on the fundamental path of the flat state and is therefore no longer the critical bifurcation point.
When $r_\mathrm{g}=3$, the local minimum for $kh_\mathrm{f}\neq0$ vanishes and the global minimum corresponds to $kh_\mathrm{f}=0$, i.e.\ sinusoidal wrinkling is no longer possible.
Briefly put, the global minimum corresponds to $kh_\mathrm{f}\neq 0$ (finite wavelength) and $kh_\mathrm{f}=0$ (infinite wavelength) before and after the Biot threshold growth ratio, respectively. At the threshold, sinusoidal wrinkling and the Biot mode are equally critical at a degenerate point of bifurcation. This transition demonstrates how wrinkling shifts from film-governed short wavelength sinusoidal wrinkling to substrate-governed infinite-wavelength Biot wrinkling when the substrate grows faster than the film.
The results from the analytical model agree well with the FE simulations up to the Biot threshold. Due to the inherent spatial dimension embedded by the finite-sized FE model, this model cannot capture the Biot mode accurately (it converges to it for very large meshes; see Supplementary Materials). The finite size of the FE model means that it converges from above to the Biot critical strain of the analytical model with increasing model dimensions (depth and width).

While the FE model cannot provide a precise quantitative prediction of the Biot critical strain in bilayers with infinite dimensions, it offers valuable qualitative insights into the mechanics underlying mode changes that occur from differential growth for bilayers of \emph{finite} dimensions. Based on the ability of the in-house FE code to pin-point bifurcation points and their evolution through parameter space~\citep{groh2018generalised,groh2022morphoelastic}, we traced the first sinusoidal wrinkling bifurcation point and its eigenmode with increasing $r_\mathrm{g}=g_\mathrm{s}/g_\mathrm{f}$. In the FE model, the boundary conditions adopted at both ends of the bilayer can only accommodate an integer number or a half-integer number of wrinkling waves and this determines the number of waves that appear in the critical eigenvector. For example, when the bilayer is modelled with width $L=4L_\mathrm{crw}$, where $L_\mathrm{crw}$ is taken from the local minimum of the analytical solution curve as shown in Figure~\ref{fig:CriticalModeChangeAnalytical}(a), the eigenmode of the first bifurcation point (the critical one) has 4 full waves, whereas the second bifurcation point has 3.5 or 4.5 full waves (see~\cite{Shen2022ProgramBilayer} for a detailed discussion).
Once the $r_\mathrm{g}$ threshold has been crossed and the Biot mode becomes critical, the critical eigenmode shifts to half of a cosine wave spanning the entire width of the bilayer. The wavelength of this cosine half-wave increases as the width of the FE model is increased (tending to infinite wavelength for infinite width) and this is how the Biot mode can be detected in an FE model of finite size. This indicates that for experimental specimens of finite size, \emph{e.g.}\ for solid blocks with $\mu_\mathrm{f}/\mu_\mathrm{s}=1$ in the mechanically compressed or constrained growth case, the Biot mode does not correspond to a solution of infinite wavelength but rather to a cosine wave of wavelength twice the the width of the specimen.  This is also why a localised crease mode is often observed in compression experiments of solid blocks; the crease mode is a post-critical evolution of the cosine half-wave eigenmode at the critical point (the Biot mode), which forms rapidly along a subcritical path away from the critical point.

Table~\ref{tab:LBAModeSeqEvolutionEfEs2.5} presents the critical growth factor, the critical nominal strain in the film and substrate and the eigenmodes of the first five bifurcation points on the fundamental path (flat state) of growing bilayers with three characteristic substrate--film growth ratios $r_\mathrm{g}$ and $\mu_\mathrm{f}/\mu_\mathrm{s}=2.5$. We start from the case with $r_\mathrm{g}=2.065$, for which the exact bifurcation condition predicts that the critical strain for sinusoidal wrinkling is equal to that of the Biot mode, see Figure~\ref{fig:CriticalModeChangeAnalytical}(b). Owing to constraints from the finite size of the FE model, the Biot wrinkling mode is less critical and therefore does not appear in the list of the first 5 eigenmodes.
There are 4, 3.5, 4.5, 3 and 5 sinusoidal waves in the eigenmodes of the first five bifurcation points.
For $r_\mathrm{g}=2.15$, there are 0.5, 4, 3.5, 4.5 and 1 sinusoidal wave(s) in the eigenmodes of the first five bifurcation points.
As described above, the one-half cosine wave corresponds to the Biot mode. However, the eigenmodes of the second to fourth bifurcation points still correspond to  sinusoidal wrinkling modes.
At $r_\mathrm{g}=2.2$, there are 0.5, 1, 1.5, 2 and 2.5 sinusoidal waves in the eigenmodes of the first five bifurcation points. In this case, all modes can be classified as Biot modes in that their wavelength increases if the model width is increased. The critical mode evolution in Table~\ref{tab:LBAModeSeqEvolutionEfEs2.5} is in accordance with the solutions of the analytical model shown in Figure~\ref{fig:CriticalModeChangeAnalytical} by demonstrating a trend of increasing wavelength with increasing $r_\mathrm{g}$, i.e.\ a rearrangement of the bifurcation point sequence from sinusoidal wrinkling governed mode (4, 3.5, 4.5, 3, 5) to a Biot governed mode (0.5, 1, 1.5, 2, 2.5).

\begin{table}[]
\centering
\caption{The eigenmodes of the first five bifurcation points on the fundamental path (flat state) of growing bilayers with different $r_\mathrm{g} = g_\mathrm{s}/g_\mathrm{f}$ ratios. Note that $\mu_\mathrm{f}/\mu_\mathrm{s}=2.5$ in all cases. The geometric and material properties of bilayers are as presented in Table~\ref{tab:bilayerSystemGeoMat}. The length of the model is $4L_\mathrm{crw}$ with the critical wrinkling wavelength $L_\mathrm{crw}$ being determined based on the local minimum with $kh\neq0$ using Eq.~\eqref{Eq:bif0}, as shown in Figure~\ref{fig:CriticalModeChangeAnalytical}; the substrate depth is $30L$.}
\label{tab:LBAModeSeqEvolutionEfEs2.5}
\begin{tabular}{cccccccccc}
\hline
$g_\mathrm{s}/g_\mathrm{f}$ & $L/h_\mathrm{f}$       & $g_\mathrm{f,cr}$   & $\varepsilon_\mathrm{f,cr}$ & $\varepsilon_\mathrm{s,cr}$ & 1st  & 2nd  & 3rd  & 4th  & 5th  \\ \hline
2.065 & 67.4701 & 0.4090 & 0.2903     & 0.4579      & 4        & 3.5      & 4.5      & 3        & 5        \\
2.15  & 81.4774 & 0.4026 & 0.2870     & 0.4640      & 0.5      & 4        & 3.5      & 4.5      & 1        \\
2.2   & 104.1449 & 0.3919 & 0.2816     & 0.4630      & 0.5      & 1        & 1.5      & 2        & 2.5      \\ \hline
\end{tabular}
\end{table}



Note that the transition from stable to unstable post-critical behaviour does not necessarily occur at the Biot threshold growth ratio that corresponds to the transition from sinusoidal wrinkling to Biot instability. Instead, this transition occurs before the Biot threshold (for smaller $g_\mathrm{s}/g_\mathrm{f}$ values) and therefore in the regime where the critical mode is of finite wavelength ($kh\neq0$). This transition is what we discuss in the next section.

\section{Initial post-buckling behaviour}
\label{Sec:PostbuckInitial}
Using the weakly nonlinear formulation of Section~\ref{Sec:theory}, we determine the initial stability of the post-buckling path branching off at the first (critical) bifurcation point. Particularly, we focus on the required combination of parameters where the initial post-buckling behaviour transitions from supercritical to subcritical.
As mentioned above, this does not always occur at the Biot threshold. 

\begin{figure}[]
\centering
\includegraphics[trim=0.0in 0in 0.0in  0in, clip=true, scale=0.7]{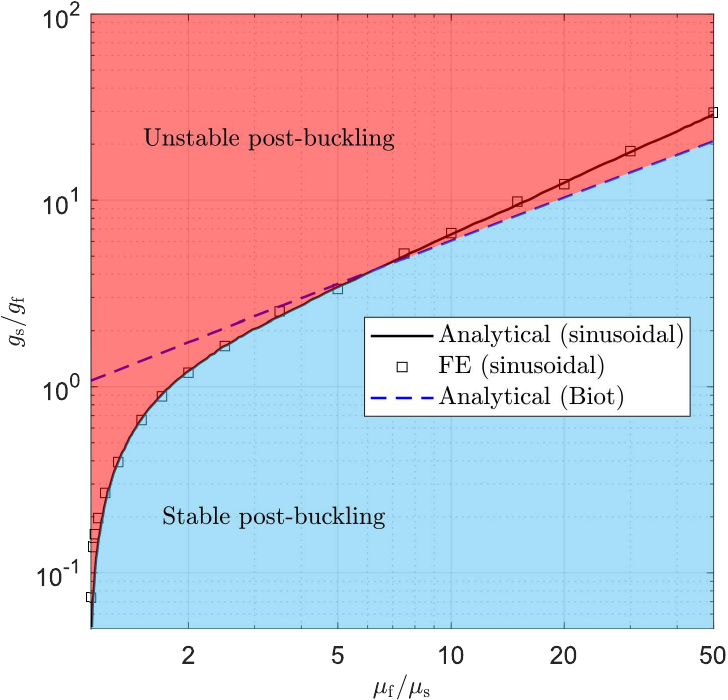}
\caption{The boundary between stable and unstable initial post-buckling as a function of  $\mu_\mathrm{f}/\mu_\mathrm{s}$ and  $g_\mathrm{s}/g_\mathrm{f}$.  The solid curve marks the supercritical to subcritical transition of the critical mode with $kh_\mathrm{f}\neq0$, whereas the dashed curve corresponds to the Biot threshold growth ratio above which the critical wavenumber becomes zero and the weakly nonlinear analysis becomes irrelevant.
%
The FE results are obtained based on an FE model with length $L=0.5L_\mathrm{crw}$ and substrate depth $d_\mathrm{s}=30L_\mathrm{crw}$.
}
\label{fig:UnstableWrinklingBCDiffGRelation}
\end{figure}

Figure~\ref{fig:UnstableWrinklingBCDiffGRelation} presents the regions of $\mu_\mathrm{f}/\mu_\mathrm{s}$ and $g_\mathrm{s}/g_\mathrm{f}$ for which the bilayer wrinkles stably or unstably and the corresponding critical buckling modes. For each $\mu_\mathrm{f}/\mu_\mathrm{s}$, the post-buckling behaviour is subcritical (unstable) if $g_\mathrm{s}/g_\mathrm{f}$ is larger than a threshold value. This threshold $g_\mathrm{s}/g_\mathrm{f}$ required to trigger unstable post-buckling behaviour increases with increasing $\mu_\mathrm{f}/\mu_\mathrm{s}$.
When the stiffness ratio $\mu_\mathrm{f}/\mu_\mathrm{s}<6.31$, the critical mode at the supercritical/subcritical threshold corresponds to sinusoidal wrinkling. However, for $\mu_\mathrm{f}/\mu_\mathrm{s}>6.31$, the critical mode transitions to the Biot mode.
When the critical mode corresponds to sinusoidal wrinkling, the analytical results compare well with our FE simulation where branch switching onto the post-buckling path and the stability of the ensuing equilibrium state is used to judge super- or subcriticality~\citep{groh2022morphoelastic}.
As mentioned in the preceding section, the FE model is not correlated to the Biot mode of the infinite half-space due to the finite size of the FE model. Therefore, only analytical solutions are presented for the Biot mode (dashed line in Figure~\ref{fig:UnstableWrinklingBCDiffGRelation}). For sinusoidal wrinkling, the correlation between analytical and FE models is excellent. Note that the weakly nonlinear analysis is conducted from the local minimum with finite wavelength rather than the global minimum with infinite wavelength (see, for example, Figure~\ref{fig:CriticalModeChangeAnalytical}(c)).


Also note that the nominal critical strain in the substrate at the transition from stable to unstable post-buckling in the regime with $\mu_\mathrm{f}/\mu_\mathrm{s}<6.31$ exceeds the experimentally observed crease strain $\varepsilon_\mathrm{crease}=0.350$~\citep{Gent1999, Hong2009, wang2015GrowthSurfaceBuckling}, but is below the Biot threshold (dashed curve).
As the corresponding critical wrinkling mode at the transitional boundary is sinusoidal with finite wavelength, this finding suggests that the unstable sinusoidal wrinkling mode may serve as a pathway to the formation of a crease mode~\citep{pandurangi2022nucleation} deeper in the post-critical regime. A fully nonlinear post-buckling analysis is required to explore this further.


\section{Advanced post-buckling behaviour and phase change diagram}
\label{Sec:AdvancePostandPhaseDiagram}
In the preceding sections, we have studied the critical and initial post-buckling behaviour of constrained growing bilayers. In this section, we will investigate the bifurcation landscape of growing bilayers under two scenarios using nonlinear FE in conjunction with continuation~\citep{groh2022morphoelastic}: scenario 1) uniform growth with varying stiffness ratios (Section~\ref{Sec:AdvancedPostUniformGrowth}), and scenario 2) differential growth rates between the film and substrate (Section~\ref{Sec:AdvancedPostDiffGrowth}). Moreover, we will discuss the onset of multi-stability in long bilayers due to differential growth (Section~\ref{Sec:AdvancedPostMultiStable}).

In the authors' previous work, \cite{Shen2022ProgramBilayer}, we have demonstrated that the substrate depth affects the bifurcation landscape of bilayers with pre-compressed substrate under end-compression, particularly in the advanced post-buckling regime. Therefore, we conducted a parametric study on the substrate depth to ensure that the depth in the FE model satisfies the infinite depth assumption. The details of this study can be seen in the Supplementary Material. We found that a depth of 30 times the critical wrinkling wavelength is sufficient for converged results. 

Previous work~\citep{Li2011,budday2015,wang2015GrowthSurfaceBuckling,Dortdivanlioglu2017} has mainly focused on the post-wrinkling behaviour of growing bilayers up to the tertiary path, i.e.\ period doubling wrinkling pattern. Here, we trace the wrinkling progress into the advanced post-wrinkling regime until creases or folds occur. Note that the full wrinkling stability landscape of growing bilayers features a large number of equilibrium paths due to successive bifurcations. For clarity and brevity, only those paths that lead to localisations at either or both boundaries are presented here.

\subsection{Uniform growth}
\label{Sec:AdvancedPostUniformGrowth}
\begin{figure}[]
\centering
\includegraphics[trim=0.0in 0.0in 0.0in 0.00in, clip=true, width=1.0\textwidth]{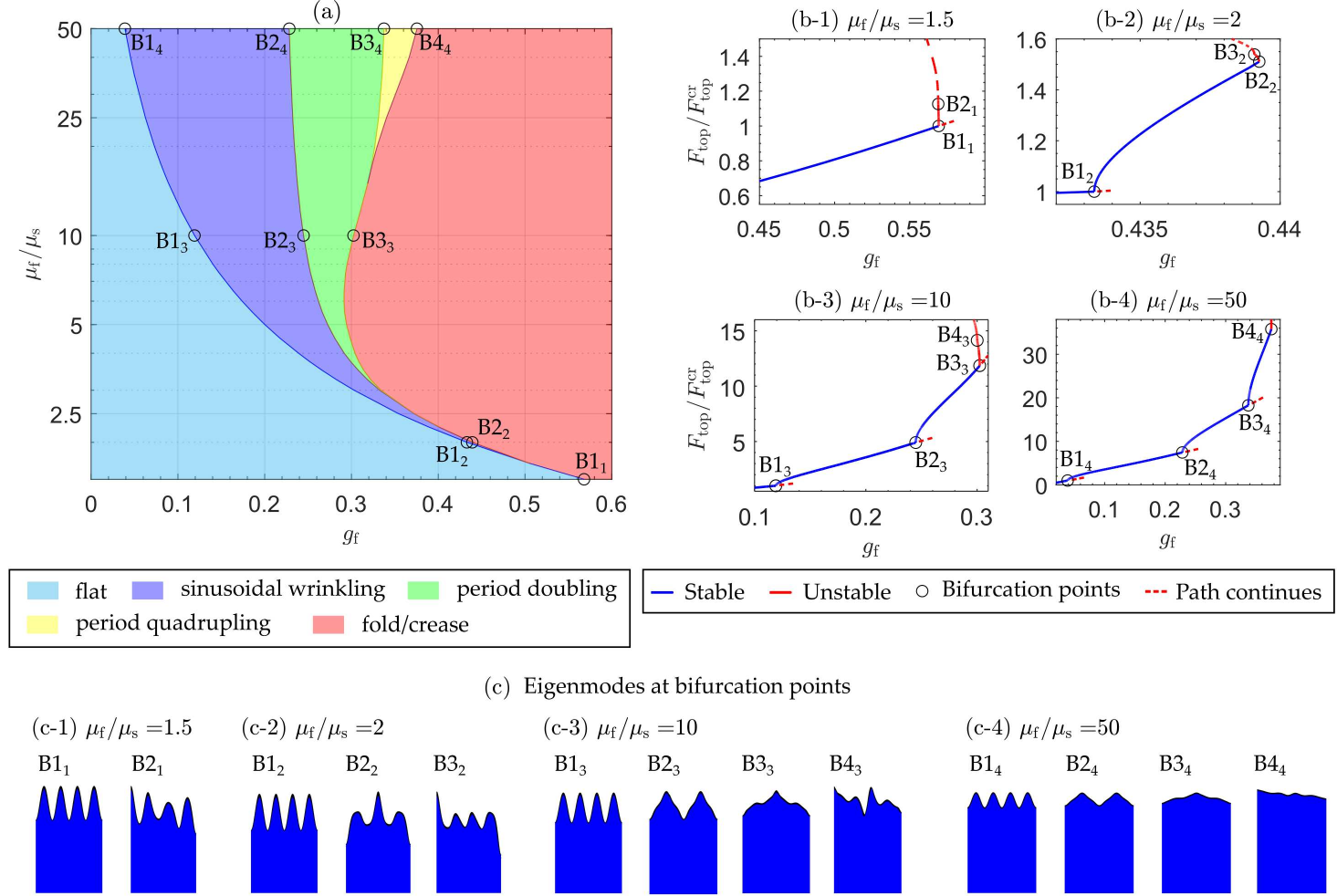}
\caption{(a) Phase change diagram of bilayers with uniform growth in the film and substrate in the regime of stiffness ratio $\mu_\mathrm{f}/\mu_\mathrm{s}=$1.5--50. Note that the phase change diagram is determined parametrically based on a finite element model with length $L=4L_\mathrm{crw}$, substrate depth $h_\mathrm{s}=30L_\mathrm{crw}$ and film thickness $h_\mathrm{f}=0.4$\ mm. The critical wrinkling wavelength $L_\mathrm{crw}$ for each case is determined using the exact bifurcation condition of Section~\ref{Sec:AnalyticalLBA}. (b) The nonlinear equilibrium path of the growing bilayers with stiffness ratios $\mu_\mathrm{f}/\mu_\mathrm{s}=$ 1.5, 2, 10 and 50, respectively. (c) The critical eigenmodes at the indicated bifurcation points. The observable bifurcation points (exchanges of stability from stable to unstable) are also labelled in the phase change diagram in (a). }
\label{fig:PhaseChangeDiagamUniformGrowth}
\end{figure}

Figure~\ref{fig:PhaseChangeDiagamUniformGrowth}(a) presents the phase change diagram for bilayers undergoing uniform growth ($g_\mathrm{s}=g_\mathrm{f}$), with stiffness ratio $\mu_\mathrm{f}/\mu_\mathrm{s}$ in the regime 1.5--50. The full equilibrium paths of bilayers with select $\mu_\mathrm{f}/\mu_\mathrm{s}$ values are presented in Figure~\ref{fig:PhaseChangeDiagamUniformGrowth}(b).
In the phase change diagram in Figure~\ref{fig:PhaseChangeDiagamUniformGrowth}(a), the different wrinkling patterns are highlighted using different colours. Their boundaries correspond to bifurcation points,  marked as hollow circles in Figure~\ref{fig:PhaseChangeDiagamUniformGrowth}(b).
Generally, for bilayers with low stiffness ratios $\mu_\mathrm{f}/\mu_\mathrm{s}$, the critical growth factor is high and these bilayers are less likely to develop advanced post-wrinkling patterns. For instance, the bilayer with $\mu_\mathrm{f}/\mu_\mathrm{s}=1.5$ does not form a stable sinusoidal wrinkling pattern despite having a sinusoidal critical eigenmode at the critical bifurcation point B1. The unstable path branching from B1 leads to another bifurcation point, B2, with an anti-symmetric critical eigenmode. Branching from B2 results in crease formation. Since all post-critical modes are unstable, these bilayers are highly sensitive to imperfections. In practice, they may jump to crease formation well below the theoretical critical growth factor. This finding suggests a potential explanation for the observed discrepancy between the critical growth factor predicted by the analytical model and the strain at which crease formation is experimentally observed. In such cases, the substrate dominates the behavior, leading to a `substrate-governed wrinkling pattern'.

With the increase of $\mu_\mathrm{f}/\mu_\mathrm{s}$, the critical growth factor decreases and the bilayer is more likely to form stable wrinkling patterns. Furthermore, the wrinkling pattern develops further into the post-critical regime. Specifically, we can observe stable sinusoidal wrinkling modes for $\mu_\mathrm{f}/\mu_\mathrm{s}>1.753$, stable period doubling modes for $\mu_\mathrm{f}/\mu_\mathrm{s}>2.5$ and stable period quadrupling modes for $\mu_\mathrm{f}/\mu_\mathrm{s}>16$. Additionally, the growth-factor regime over which these wrinkling patterns remain stable generally increases with $\mu_\mathrm{f}/\mu_\mathrm{s}$. In these cases, where sinusoidal wrinkling, period doubling and period quadrupling precede crease formation, we may define the wrinkling pattern as the `film-governed wrinkling mode'.
The pathway to crease formation differs depending on the stiffness ratio.
For $\mu_\mathrm{f}/\mu_\mathrm{s}=2$, as shown in Figure~\ref{fig:PhaseChangeDiagamUniformGrowth}(b-2), the stable sinusoidal wrinkling pattern loses stability at B2 with the critical eigenmode being period quadrupling. Bifurcation from B2 leads to an unstable path with another bifurcation point B3, where the critical eigenmode is then anti-symmetric. Bifurcation from B3 leads to unstable path and the formation of a crease.
For $\mu_\mathrm{f}/\mu_\mathrm{s}=10$, see Figure~\ref{fig:PhaseChangeDiagamUniformGrowth}(b-3), the bilayer can form stable sinusoidal and period doubling patterns. Beyond bifurcation point B3, the corresponding period doubling mode becomes unstable and the subsequent pathway to creasing is similar to that in $\mu_\mathrm{f}/\mu_\mathrm{s}=2$, where there are 2 sequential bifurcation points on the unstable post-buckling path with the critical eignmodes being period quadrupling and of anti-symmetric shape.
For $\mu_\mathrm{f}/\mu_\mathrm{s}=50$, the doubling in wrinkling period is stable up to period quadrupling until the bilayer becomes unstable at bifurcation point B4, with the critical eigenmode again being anti-symmetric. Note that all the eigenmodes at B1--B4 are essentially identical to those in $\mu_\mathrm{f}/\mu_\mathrm{s}=10$. The only difference is that the path bifurcated from B3 is stable in this case. Hence, in all cases, the crease forms via an unstable bifurcated path from a critical point with an anti-symmetric critical eigenvector.

\subsection{Differential growth}
\label{Sec:AdvancedPostDiffGrowth}
In this section, we explore how the differential growth in film and substrate affects the wrinkling behaviour of growing bilayers, where film and substrate grow in an isotropic pattern but at different rates ($g_\mathrm{s}\neq g_\mathrm{f}$). Here, we particularly focus on two characteristic  stiffness ratios $\mu_\mathrm{f}/\mu_\mathrm{s}=10$ and $2.5$. The  growth ratio $g_\mathrm{s}/g_\mathrm{f}$ varies from 0 to the transitional state where the critical mode becomes the Biot mode, as shown in Figure~\ref{fig:UnstableWrinklingBCDiffGRelation}. This range is significantly wider compared to previous studies~\citep{Dortdivanlioglu2017}, where the growth ratio $g_\mathrm{s}/g_\mathrm{f}$ is limited to 0 -- 1. By exploring a broader growth ratio range, we gain a deeper insight into the bilayer's growth mechanics.

\begin{figure}[]
\centering
\includegraphics[trim=0.0in 0.0in 0.0in 0.00in, clip=true, width=1.0\textwidth]{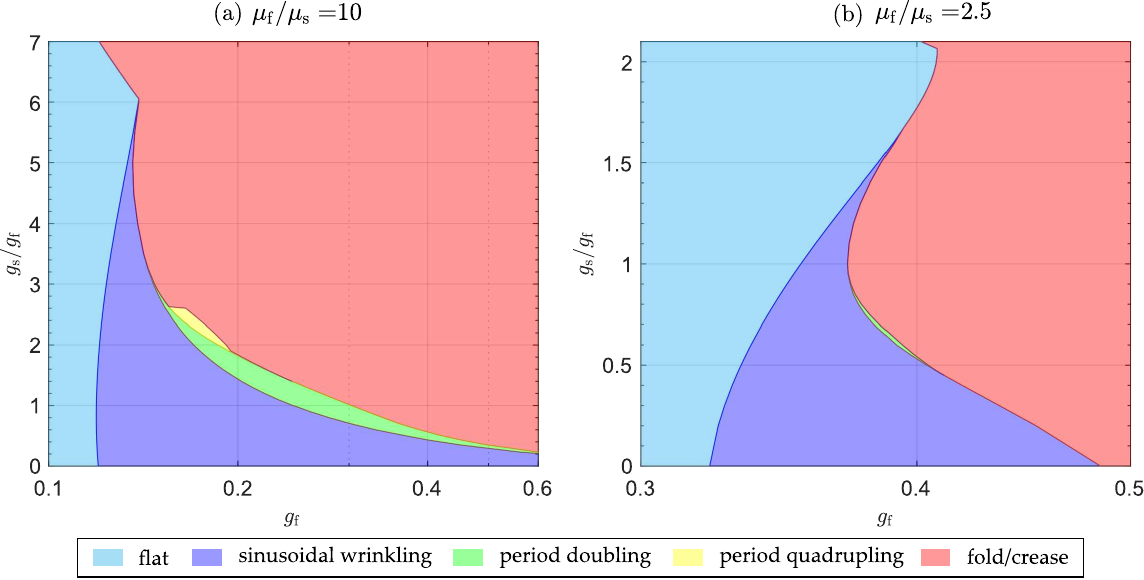}
\caption{Phase change diagram of bilayers with differential growth in the film and substrate with stiffness ratios of $\mu_\mathrm{f}/\mu_\mathrm{s}=10$ and $\mu_\mathrm{f}/\mu_\mathrm{s}=2.5$.
The details are as described in Figure~\ref{fig:PhaseChangeDiagamUniformGrowth}(a). The equilibrium paths of the bilayers at selected $g_\mathrm{s}/g_\mathrm{f}$ ratios can be found in the Supplementary Materials. }
\label{fig:PhaseChangeDiagamDiffGrowth}
\end{figure}

Figure~\ref{fig:PhaseChangeDiagamDiffGrowth} presents the phase change diagram of bilayers with $\mu_\mathrm{f}/\mu_\mathrm{s}=10$ or 2.5. Generally, with the increase of $g_\mathrm{s}/g_\mathrm{s}$, the critical mode transitions from a `film-governed' wrinkling mode to `substrate-governed' wrinkling mode. Specifically, the regime for the stable wrinkling pattern becomes narrower. The kink on the boundary between the flat (blue) and sinusoidal wrinkling (purple) and crease (red) corresponds to the state where the Biot mode becomes the critical mode, which is identical to the kinks seen in the subfigures of the first column of Figure~\ref{fig:fig4}. The parameters for which these kinks occur are determined using the analytical model based on the condition that the local minimum at $kh_\mathrm{f}\neq0$ is identical to that for $kh_\mathrm{f}=0$, as illustrated in Figure~\ref{fig:CriticalModeChangeAnalytical} for $\mu_\mathrm{f}/\mu_\mathrm{s}=2.5$. The values of $g_\mathrm{f}/g_\mathrm{s}$ at the kink for $\mu_\mathrm{f}/\mu_\mathrm{s}=10$ and 2.5 are $g_\mathrm{f}/g_\mathrm{s}=6.055$ and 2.065, respectively. All data for $g_\mathrm{f}/g_\mathrm{s}$ values above the kink are from the analytical model. Our findings regarding the critical growth factor differ from a previous study~\citep{Dortdivanlioglu2017} where the authors concluded that the critical growth factor is independent of the growth ratio. Instead, our results suggest that this is not the case when the stiffness ratio $\mu_\mathrm{f}/\mu_\mathrm{s} \rightarrow 1$ and/or the substrate grows  faster than the film (i.e. $g_\mathrm{s}/g_\mathrm{f}>1$).

An increase of $g_\mathrm{s}/g_\mathrm{f}$ generally decreases the regime of stable sinusoidal wrinkling, but has more complicated effects on the advanced wrinkling regime. For both cases of $\mu_\mathrm{f}/\mu_\mathrm{s}=2.5$ and $\mu_\mathrm{f}/\mu_\mathrm{s}=10$, when $g_\mathrm{s}/g_\mathrm{f}$ approaches 0, the stable period doubling mode disappears. As reported by previous studies~\citep{liu2017robust,Dortdivanlioglu2017}, the nonlinearity in the substrate arising from pre-strain or compression is critical for the formation of the period doubling mode. Interestingly, for $\mu_\mathrm{f}/\mu_\mathrm{s}=10$, in the regime of $g_\mathrm{s}/g_\mathrm{f}=[1.5 2.625]$, there also exists stable period quadrupling, which is not observed in the case with uniform growth. However, this is not observed for $\mu_\mathrm{f}/\mu_\mathrm{s}=2.5$.

The interplay between film-to-substrate stiffness ratio $\mu_\mathrm{f}/\mu_\mathrm{s}$ and growth ratio $g_\mathrm{s}/g_\mathrm{f}$ dictates the formation and evolution of wrinkling patterns in growing bilayers. These two factors act as competing influences, shaping the resulting wrinkle complexity. For very high $g_\mathrm{s}/g_\mathrm{f}$ ratios the substrate dominates the wrinkling behaviour such that the first instability is always into the Biot mode. The same is true for uniform growth ($g_\mathrm{s}/g_\mathrm{f}=1$) when the film and substrate have similar elastic moduli. For very low $g_\mathrm{s}/g_\mathrm{f}$ ratios and similar film and substrate moduli, the substrate does not present a sufficiently pronounced nonlinear foundation for the film to develop period doubling bifurcations. Hence, for period-doubling bifurcations to occur, the film has to be of the order of 5 to 10 times stiffer than the substrate and the substrate cannot grow too quickly and neither too slowly compared to the film. This also explains why it is not possible to introduce a unified equivalent stiffness ratio to describe the phase change diagram of bilayers under differential growth, based on the uniform growing case, as the behaviour of differentially growing bilayers is qualitatively different.

\subsection{Multiple stable post-buckling states due to faster growth in the substrate}
\label{Sec:AdvancedPostMultiStable}
In the previous sections, phase change diagrams were generated using bilayers with a length of $L = 4L_\mathrm{crw}$. However, in previous research involving end-compressed bilayers with highly pre-compressed substrates~\citep{Shen2022ProgramBilayer}, we identified the presence of multi-stable post-buckling modes in the intermediate post-buckling regime, known as `building blocks'. Specifically, after bifurcating from a flat into a sinusoidally wrinkled state, bilayers undergo further $n$-tupling bifurcations into stable wrinkling patterns of longer wavelength. The periodicity of these patterns, denoted as $n$, can range from 4 to 8 and is determined by the overall length of the bilayer and the pre-compression strain in the substrate.

\begin{figure}[]
\centering
\includegraphics[trim=0.0in 0.0in 0.0in 0.00in, clip=true, width=1.0\textwidth]{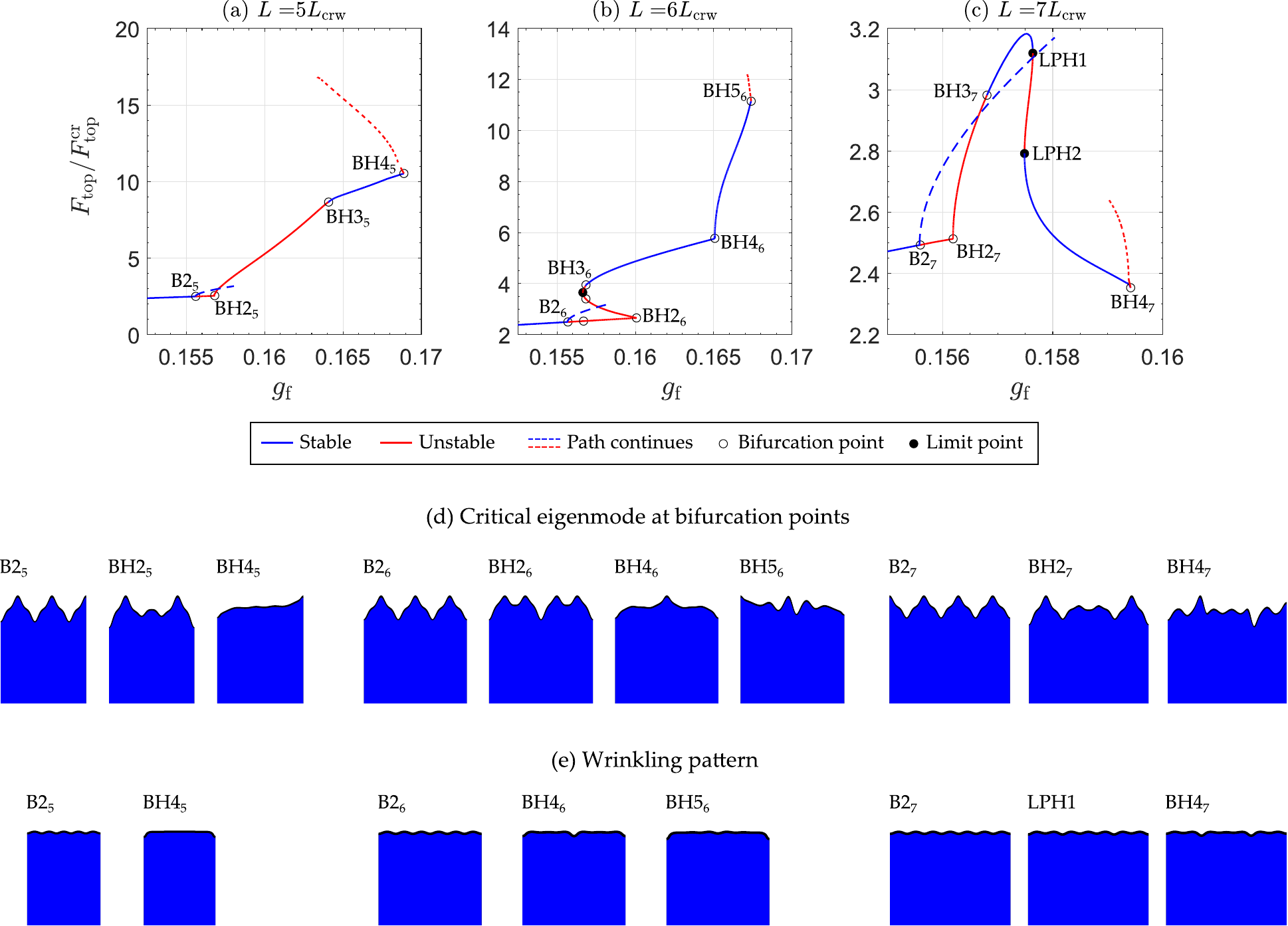}
\caption{Equilibrium path and wrinkling patterns of growing bilayers with $\mu_\mathrm{f}/\mu_\mathrm{s}=10$ and $g_\mathrm{s}/g_\mathrm{f}=2.5$ for growing bilayers with (a) $L=5L_\mathrm{crw}$, (b) $L=6L_\mathrm{crw}$, (c) $L=7L_\mathrm{crw}$. The details of the graphs are as described in Figure~\ref{fig:PhaseChangeDiagamUniformGrowth}(b).  
Note that the fundamental paths are not shown for clarity. }
\label{fig:fig9}
\end{figure}

Here, we also observe the existence of stable $n$-tupling modes within the intermediate post-buckling regime for cases where the substrate grows faster than the film. Figure~\ref{fig:fig9} illustrates the equilibrium path and wrinkling patterns of growing bilayers with parameters $\mu_\mathrm{f}/\mu_\mathrm{s} = 10$ and $g_\mathrm{s}/g_\mathrm{f} = 2.5$, and lengths $L = 5L_\mathrm{crw}$, $6L_\mathrm{crw}$ and $7L_\mathrm{crw}$. For the bilayer with $L=5L_\mathrm{crw}$, we observe stable period quintupling, see BH3$_5$--BH4$_5$ in Figure~\ref{fig:fig9}(a). For the bilayer with $L=6L_\mathrm{crw}$, we observe stable period tripling, see BH3$_6$--BH4$_6$ in Figure~\ref{fig:fig9}(b), and its further doubling leads to stable period sextupling, BH4$_6$--BH5$_6$ in Figure~\ref{fig:fig9}(b). For the bilayer with $L=7L_\mathrm{crw}$, stable period septupling is observed. As the growth factor increases further, the wrinkling pattern evolves into a combination of adjacent stable period quadrupling and period tripling modes. Therefore, we conclude that the stable building blocks for the bespoke bilayer are period tripling, period quadrupling and period quintupling. Unlike end-compressed bilayers with pre-compressed substrates, growing bilayers with a fixed $g_\mathrm{s}/g_\mathrm{f}$ ratio tend to develop creases through unstable bifurcations rather than by achieving stable period doubling to form folds. This behavior is attributed to the increasing dominance of the substrate.
It is important to note that these stable wrinkling patterns are separated from the natural loading path by segments of unstable equilibria. Therefore, the progression of wrinkling patterns from flat to sinusoidal wrinkling, followed by doubling in period, is preferred if there are no or small imperfections.

However, the presence of random imperfections with sufficient magnitude can lead to irregular wrinkling patterns~\citep{auguste2014PostWringBifur}. In such cases, the actual pattern formation becomes a random combination of different building blocks governed by these imperfections~\citep{Shen2022ProgramBilayer}. To address this challenge, we could adopt \emph{modal nudging} as a means to tailor the pattern formation, as discussed in our previous works \citep{COX2018Nudge,Shen2022ProgramBilayer,Shen2023ActiveNuding}. Specifically, the deformation modes affine to one of the stable post-critical equilibria, $n$-tupling modes, are introduced as geometric perturbation to alter the undeformed baseline geometry of the bilayer, such that the bilayer favours the seeded post-buckling response over potential alternatives. The size and pattern of these perturbation are important for the effectiveness of this technique. The size must be large enough to overcome or erode the energy barrier between the desired isolated equilibria and the natural loading path~\citep{COX2018Nudge}. In terms of the pattern, our previous work has shown that localizing dents at positions where creases form can effectively achieve the desired outcome \citep{Shen2022ProgramBilayer}. For $n$-tupling modes with longer characteristic wavelength, additional dents may be required to guarantee \emph{nudging authority}~\citep{SHEN2021ExpPathFollowPart2}. More details can be found in \cite{Shen2022ProgramBilayer}.
\section{Concluding remarks}
\label{Sec:Conclu}
Through analytical modelling and nonlinear FE simulation, we have provided a complete understanding of the nonlinear mechanics governing growing bilayers in a regime which has previously not been explored in detail, i.e.\ similar film and substrate moduli and a substrate that grows faster than the film. The high nonlinearity, transition to subcriticality and the onset of the Biot mode in these scenarios implies a qualitatively different behaviour from the previously studied cases where the stiffness ratio is large and only the film grows~\citep{budday2015,Dortdivanlioglu2017,alawiye2019revisiting,Alawiye2020}.

We have shown that the commonly employed first-order asymptotic expansion of the nominal critical strain for uniform growth (or uniform mechanical compression) loses accuracy for $\mu_\mathrm{f}/\mu_\mathrm{s}<50$. We provided higher-order correction terms so as to accurately describe the critical wrinkling states in the specific parameter regimes. There exists a particular combination of $\mu_\mathrm{f}/\mu_\mathrm{s}$ and $g_\mathrm{s}/g_\mathrm{f}$ for which wrinkling transitions from super- to subcritical and this occurs for sinusoidal wrinkling of finite wavelength rather than Biot wrinkling of infinite wavelength. The Biot wrinkling threshold occurs for a separate locus of $\mu_\mathrm{f}/\mu_\mathrm{s}$ and $g_\mathrm{s}/g_\mathrm{f}$ pairs.


Beyond the onset of wrinkling, the post-buckling regime can be split into regions of sinusoidal wrinkling, period doubling, period quadrupling and creasing which are all or partially traversed with increasing growth magnitude depending on the combination of $\mu_\mathrm{f}/\mu_\mathrm{s}$ and $g_\mathrm{s}/g_\mathrm{f}$. Our analysis revealed two key factors that compete to influence the formation of advanced wrinkling patterns: (i) the nonlinear behavior of the substrate arising from compression strain and (ii) the film-to-substrate stiffness ratio.  Moreover, we have shown that the formation of the crease occurs along a post-buckling path seeded by an anti-symmetric eigenvector and not through a direct bifurcation off the fundamental path as has been previously surmised~\citep{hohlfeld2011unfolding}. Finally, we have demonstrated the existence of multi-stability in the advanced post-buckling regimes for growing bilayers where growth in the substrate surpasses that of the film.

As for the identified size effects of bilayer models, while it is convenient to assume an infinite bilayer in the analytical model, in reality, biological bilayers always have finite size. One implication of this is that Biot wrinkling eigenmode now takes the form of a half-cosine wave along the model length, with the wavelength therefore scaling linearly with model length to reach the asymptotic limit of infinite wavelength. Additional constraints resulting from the finite size of the model must be considered to predict their response accurately. In these cases, the FE model may have advantages over analytical modelling. In previous work~\citep{Shen2022ProgramBilayer}, we have demonstrated that the substrate depth can have a strong qualitative effect on the bifurcation landscape. Therefore, an appropriate choice of model size that reflects the physical specimen and analysis purpose is important for accurate and reliable modelling.
Our results are presented using growth factor, which allows for the calculation of nominal strain in both film and substrate. Based on the findings of \cite{HOLLAND2017SoftFilm} regarding the equivalence of nominal strain arising from both mechanical loading and growth, the mechanics we have unveiled are also applicable to other loading scenarios, such as mechanical (pre-stretch)~\citep{AUGUSTE2017PostWrinkleExp} and active materials~\citep{agrawal2014shape}. Our results therefore facilitate an understanding of the nonlinear behaviour of bilayers with small stiffness ratios and highly pre-compressed substrates in many applications of surface texturing~\citep{yoo2003evolution} and the manufacturing of flexible electronics~\citep{khang2006stretchable}.

\section*{Author contributions}
\textbf{JS}: Conceptualization, Data curation, Formal analysis, Investigation, Methodology, Validation, Visualization, Writing - original draft;
\textbf{YF}: Conceptualization, Data curation, Formal analysis, Investigation, Methodology, Validation, Resources, Supervision, Funding acquisition, Visualization, Writing - original draft, review\&editing;
\textbf{AP}: Conceptualization, Resources, Supervision, Project administration, Writing - review\&editing;
\textbf{RG}: Conceptualization, Investigation, Methodology, Project administration, Resources, Software, Supervision, Funding acquisition, Writing - review\&editing.

\section*{Acknowledgements}
This work was funded by the Leverhulme Trust through a Philip Leverhulme Prize awarded to RMJG. RMJG was also funded by the Royal Academy of Engineering under the Research Fellowship scheme [RF$\setminus$201718$\setminus$17178]. YF was funded by the Engineering and Physical Sciences Research Council, UK (Grant No EP/W007150/1).


\end{document}